\documentclass[prb,amsmath,amssymb,twocolumn]{revtex4-1}
\usepackage{graphicx}
\usepackage{subfigure}
\usepackage{adjustbox}
\usepackage{bm}
\usepackage{color}
\usepackage{braket}
\usepackage{standalone}
\usepackage{multirow}
\usepackage{tikz}
\usepackage{mathrsfs}
\usepackage{dsfont}

\begin{document}

\title{Unconventional $\mathbb{Z}_{n}$ parton states at $\nu = 7/3$: The role of finite width}
\author{William N. Faugno$^{1}$, Tongzhou Zhao$^{1}$, Ajit C. Balram$^{2}$, Thierry Jolicoeur$^{3}$, and Jainendra K. Jain$^{1}$}
\affiliation{$^{1}$Department of Physics, 104 Davey Lab, Pennsylvania State University, University Park, Pennsylvania 16802, USA}
\affiliation{$^{2}$Institute of Mathematical Sciences, HBNI, CIT Campus, Chennai 600113, India}
\affiliation{$^{3}$Institut de Physique Th\'eorique, Universit\'e Paris-Saclay, CNRS, CEA, 91190 Gif sur Yvette, France}
\date{\today}

\begin{abstract} 
A recent work [Balram, Jain, and Barkeshli, Phys. Rev. Res. ${\bf 2}$, 013349 (2020)] has suggested that an unconventional state describing $\mathbb{Z}_{n}$ superconductivity of composite bosons, which supports excitations with charge $1/(3n)$ of the electron charge, is energetically better than the Laughlin wave function at $\nu=7/3$ in GaAs systems. All experiments to date, however, are consistent with the latter. To address this discrepancy, we study the effect of finite width on the ground state and predict a phase transition from an unconventional $\mathbb{Z}_{n}$ state at small widths to the Laughlin state for widths exceeding $\sim$ 1.5 magnetic lengths. We also determine the parameter region where an unconventional state is stabilized in the one third filled zeroth Landau level in bilayer graphene. The roles of Landau level mixing and spin are also considered.
\end{abstract}

\maketitle

\section{Introduction}
\label{sec:Intro}

The fractional quantum Hall effect (FQHE) at filling factor $\nu = 1/3$ in the lowest Landau level (LLL) was the first to be experimentally observed~\cite{Tsui82} and subsequently understood, as a result of Laughlin's trial wave function~\cite{Laughlin83}. The FQHE has since proven to be an incredibly rich platform for exploring the physics of strongly correlated electron systems. After the initial observation, many additional fractions were observed in the LLL at $\nu = n/(2pn\pm1)$ with $n$ and $p$ positive integers. These fractions are understood as an integer quantum Hall effect (IQHE) of composite fermions, where a composite fermion (CF) is an emergent particle consisting of an electron bound to an even number of vortices~\cite{Jain89, Jain07, Halperin20}. 

In contrast, the FQHE in the second LL (SLL) of GaAs is less well understood. Interestingly, even the physical origin of the $\nu=7/3$ FQHE, which corresponds to 1/3 filled SLL, has not been conclusively established. Exact diagonalization studies~\cite{Ambrumenil88, Peterson08, Peterson08b, Balram13b, Kusmierz18, Balram20} have convincingly shown that the actual state at $\nu=7/3$ for a zero width system is an incompressible FQHE state. However, the overlap of the exact ground state with the Laughlin state is not large, typically less than $60\%$ for systems accessible to exact diagonalization studies~\cite{Ambrumenil88, Peterson08, Peterson08b, Balram13b, Kusmierz18, Balram20}. [In contrast, the overlap of the Coulomb ground state at $\nu=1/3$ in the LLL with the Laughlin state is greater than $98\%$ for up to $N=15$ electrons~\cite{Balram20}.] Furthermore, the excitations of the 7/3 FQHE in exact diagonalization studies are qualitatively different from those at 1/3, and exact diagonalization studies also do not show a clearly identifiable branch of low-energy excitations, called the magnetoroton or the CF-exciton mode, as observed in the LLL~\cite{Girvin85, Girvin86, Dev92, Scarola00, Jolicoeur17}. As a result, the precise nature of the state at $\nu=7/3$ has remained a topic of debate~\cite{Read99, Balram13b, Johri14, Zaletel15, Peterson15, Jeong16, Balram19a, Balram20}. 

In a recent work, Balram \emph{et al}.\cite{Balram19a} have proposed, inspired by the parton paradigm for the FQHE~\cite{Jain89b}, that the $\nu=7/3$ FQHE is a $\mathbb{Z}_n$ topological superconductor, wherein bound states of $n$ composite bosons~\cite{Zhang89} undergo Bose-Einstein condensation. This generalizes the Zhang-Hansson-Kivelson theory of the 1/3 Laughlin state as a Bose-Einstein condensate of composite bosons~\cite{Zhang89}, with $\mathbb{Z}_1$ corresponding to the Laughlin wave function. While the different $\mathbb{Z}_n$ states share many topological quantum numbers, a key distinction between them is that the elementary quasiparticle has a charge of $-e/(3n)$, where $-e$ is the charge of the electron. Variational calculations in Ref.~\cite{Balram19a} suggest that the best candidate is the $\mathbb{Z}_3$ state, which has lower energy than the Laughlin state in the thermodynamic limit, and also a higher overlap with the exact SLL Coulomb ground state for systems where such a calculation is possible.

The experimental observations at $\nu = 7/3$ are, however, largely consistent with the Laughlin state. In particular, shot noise~\cite{Dolev08, Venkatachalam11,Dolev11} and scanning single-electron transistor~\cite{Venkatachalam11} experiments at $\nu = 7/3$ have measured quasiparticles of charge $-e/3$. This raises the question: Why are experimental measurements consistent with the Laughlin state while theory suggests that better variational states exist? This question has motivated the present study.

There can be several reasons for the discrepancy between theory and experiment. The theoretical calculations mentioned above do not include the effects of finite width, Landau level mixing, screening, and disorder, which can affect the variational comparisons. We consider in this article the competition between the different $\mathbb{Z}_n$ states as a function of the quantum well width.  Our primary result is the prediction of a phase transition from the $\mathbb{Z}_4$ state at small widths into the Laughlin ($\mathbb{Z}_1$) state when the quantum well width exceeds approximately 1.5 magnetic lengths. We also predict a similar phase transition at $\nu=1/3$ in the zeroth Landau level of bilayer graphene as a function of the magnetic field.

\section{$\mathbb{Z}_n$ parton wave function at $\nu=1/3$}
\label{sec:wftns}
The parton theory generalizes the Jain CF states~\cite{Jain07} to a larger class of candidate wave functions~\cite{Jain89b}. In the parton theory, one considers fractionalizing electrons into a set of fictitious particles called partons. The partons are fractionally charged, have the same density as electrons, and have filling factor $\nu_\alpha$, where $\alpha$ labels the parton species.  An incompressible state is achieved when each parton species is in an IQHE state, i.e. $\nu_\alpha = n_\alpha$, with $n_\alpha$ an integer. (More generally, we can place the partons in any known incompressible states.) The partons are of course unphysical and must be combined back into physical electrons, which is equivalent to setting the parton coordinates $z_j^\alpha$ equal to the parent electron coordinates $z_j$, i.e. $z_j^\alpha = z_j$ for all $\alpha$. (The quantity $z_j = x_j - iy_j$ is the complex coordinate of the $j$th electron.) The resulting wave functions, labeled ``$n_1n_2n_3...$," are given by
\begin{equation}
\Psi^{n_1n_2n_3...}_\nu = \mathcal{P}_{\rm LLL} \prod_{n_\alpha} \Phi_{n_\alpha}(\{z_j\}),
\end{equation}
where $\Phi_n$ is the Slater determinant wave function for the state with $n$ filled Landau levels, and $\mathcal{P}_{\rm LLL}$ denotes projection into the LLL, as appropriate in the high field limit. The partons can also experience magnetic fields anti-parallel to the field experienced by electrons; these correspond to negative filling factors, which we denote as $\bar{n}$, with $\Phi_{\bar{n}}=\Phi_{-n}=\Phi_n^*$. To ensure that each parton species has the same density as the electron density, the charge of each parton species is given by $e_\alpha = -\nu e / \nu_\alpha$. The relation $\sum_\alpha e_\alpha=-e$ implies that the electron filling factor is given by $\nu = [\sum_\alpha \nu_\alpha^{-1}]^{-1}$. The Laughlin wave function at $\nu=1/3$ can be interpreted as the $111$ parton state. The Jain $n/(2pn+1)$ states appear as the $n11...$ states and the Jain $n/(2pn-1)$ states as $\bar{n}11...$; these correspond to the wave function $\Psi_{n/(2pn\pm1)} = \mathcal{P}_{\rm LLL} \Phi_{\pm n} \Phi_1^{2p}$. Many other parton states have recently been shown to be plausible for SLL and other FQHE~\cite{Wu17, Balram18, Bandyopadhyay18, Balram18a, Faugno19, Balram19, Kim19, Balram19a, Balram20, Faugno20a, Balram20a}. These states often have exotic properties, such as non-Abelian anyonic excitations~\cite{Wen91}.  

For $\nu = 1/3$, Balram \emph{et al}. proposed the $\mathbb{Z}_n$ parton states described by the wave function
\begin{equation}
\Psi_{1/3}^{\mathbb{Z}_n} = \mathcal{P}_{\rm LLL}\Phi_n\Phi_{\bar{n}}\Phi_1^3 \sim \Psi_{n/(2n+1)}\Psi_{n/(2n-1)} \Phi_1^{-1},
\end{equation}
where in the last step we redefine the wave function as $[\mathcal{P}_{\rm LLL}\Phi_n \Phi_1^2] [\mathcal{P}_{\rm LLL}\Phi_{\bar{n}} \Phi_1^2]\Phi_1^{-1}$. (This grouping is chosen to facilitate the use of Jain-Kamilla projection method in our numerical simulations \cite{Jain07,Jain97, Jain97b, Davenport12, Moller05, Balram15a}. It is accepted and has been shown for many cases that the topological properties of the state do not depend on the details of the projection method \cite{Balram16b}.) Because the factor $\Phi_n\Phi_{\bar{n}}$ is real, all the $\mathbb{Z}_n$ states occur at the same ``shift"~\cite{Wen92} $\mathcal{S}=3$ in the spherical geometry. The physical interpretation of the wave function as a superconductor of composite bosons arises from the fact that $\Phi_n\Phi_{\bar{n}}$ represents a $\mathbb{Z}_n$ superconductor of electrons~\cite{Barkeshli13,Balram19a}, and the factor $\Phi_1^3$ attaches three vortices to each electron to convert it into a composite boson. The elementary excitation corresponds to an excitation in the factor $\Phi_n$ or $\Phi_{\bar{n}}$ and has a charge of magnitude $e/(3n)$. 

\section{Finite width phase diagram for $\nu=7/3$ in GaAs}

We will use the spherical geometry~\cite{Haldane83} in our calculations, which considers $N$ electrons on the surface of a sphere subjected to a total flux $2Q\phi_0$, with $\phi_0=hc/e$ and $2Q$ is a positive integer. The radius of the sphere is $R=\sqrt{Q}\ell$, where $\ell=\sqrt{\hbar c/eB}$ is the magnetic length. The state with $n$ filled Landau levels can only be constructed for particle number $N$ divisible by $n$ and $N\geq n^2$. The same is true of the $\mathbb{Z}_n$ state.

We approximate the confining potential as an infinite square well of width $w$ that results in a transverse wave function given by a sine function. The problem of electrons in the SLL interacting with the Coulomb interaction is equivalent to that of electrons in the LLL interacting with an effective interaction; the effective interaction for this system is given in Ref.~\cite{Toke08}. We consider well widths up to 5 magnetic lengths. (For convenience, we use the disk pseudopotentials for our calculations in the spherical geometry; this should not cause any corrections because we will perform the calculation for large systems and take the thermodynamic limit.)

\begin{figure}[tbhp]
\includegraphics[width=0.49\textwidth]{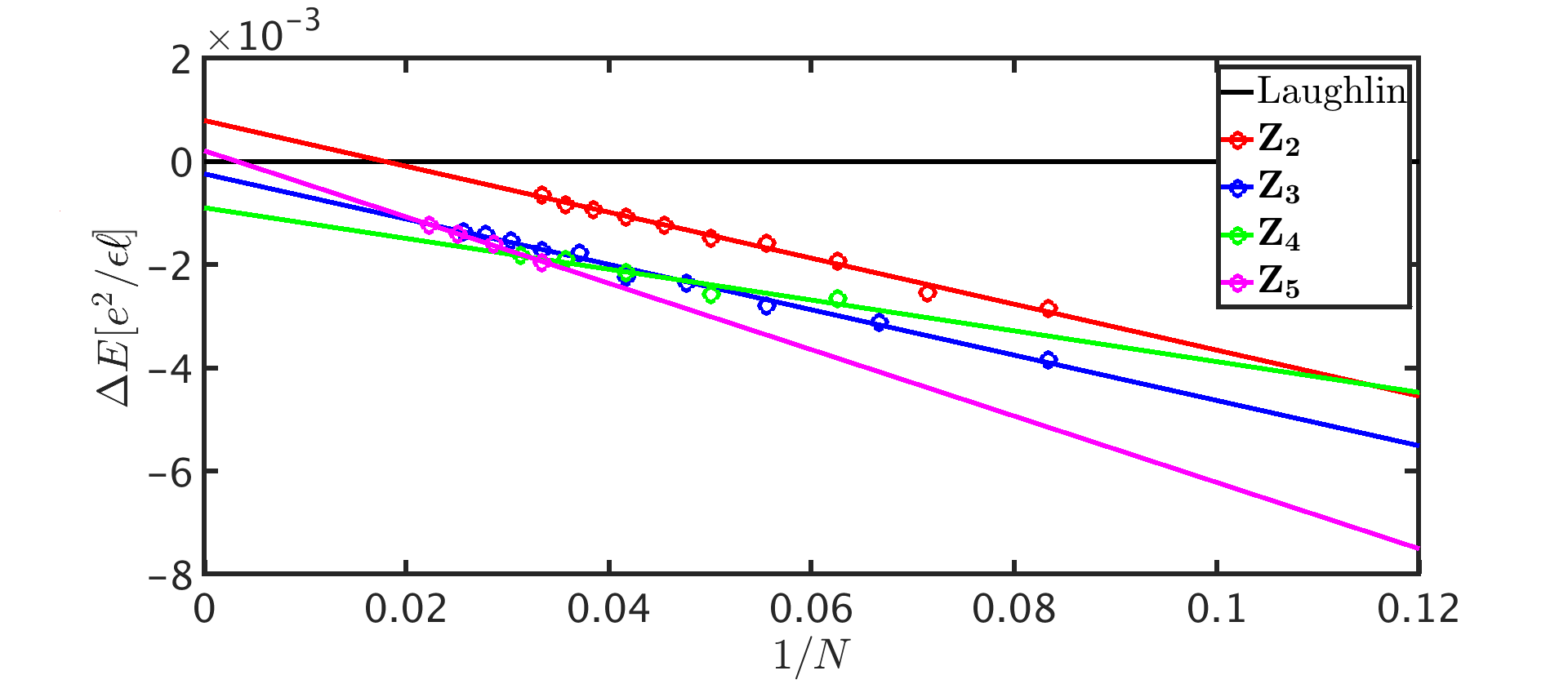}
\caption{Thermodynamic extrapolations for the energy differences $\Delta E=E(\mathbb{Z}_n)-E_{\rm Laughlin}$ for various $\mathbb{Z}_n$ states at $\nu=7/3$. The results are for zero quantum well width. The $\mathbb{Z}_4$ state has the lowest energy in the thermodynamic limit.}
\label{fig: TLs}
\end{figure}

\begin{figure}[tbhp]
\includegraphics[width=0.49\textwidth]{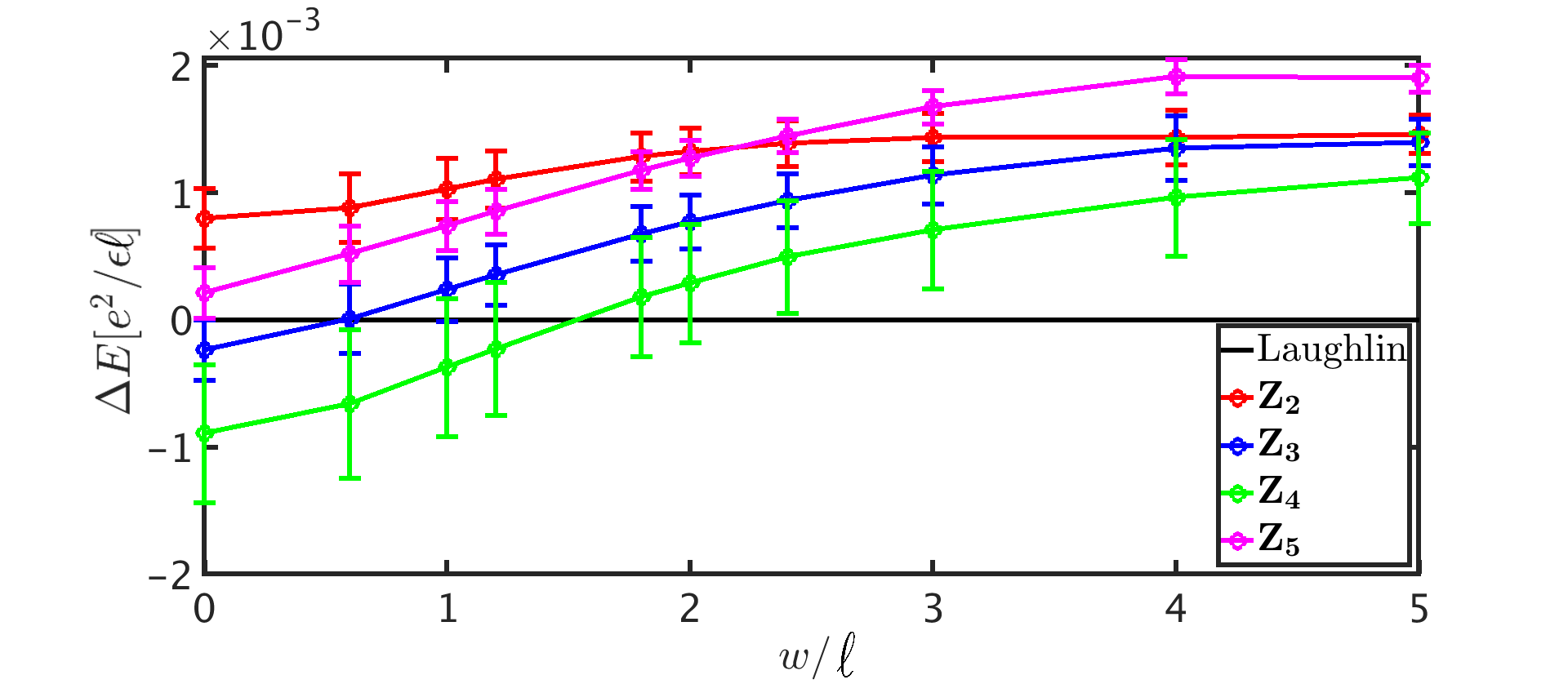}
\caption{Thermodynamic energies of various $\mathbb{Z}_n$ states at $\nu=7/3$ as a function of the quantum well width $w$. All energies are quoted relative to the Laughlin state.  The $x$-axis is the quantum well width in units of magnetic length. A phase transition from the $\mathbb{Z}_4$ to the Laughlin state is seen to occur at $w\sim1.5 \ell$.}
\label{fig: FWSLL}
\end{figure}

At zero width, the $\mathbb{Z}_4$ state has the lowest energy in the thermodynamic limit, as seen in Fig.~\ref{fig: TLs}. Here and below, all energies are quoted in units of $e^2/(\epsilon \ell)$ where $\epsilon$ is the dielectric constant of the material. (We note that the $\mathbb{Z}_4$ and $\mathbb{Z}_5$ states were not studied in Ref.~\cite{Balram19a}. Also, we cannot definitively rule out the $\mathbb{Z}_3$ state within the numerical accuracy of our calculations.) We have similarly determined the thermodynamic energies for quantum wells of various widths [see Appendix~\ref{sec: ThermoLim}]. In Fig.~\ref{fig: FWSLL}, we show how the energies of several $\mathbb{Z}_n$ states, measured relative to the energy of the Laughlin state, evolve as we increase the well-width. For a $w<1.5\ell$, the $\mathbb{Z}_4$ state has the lowest variational energy in the thermodynamic limit. For $w>1.5\ell$, the Laughlin state is preferred in our calculation, suggesting that the $\mathbb{Z}_4$ state should only be observed in samples with sufficiently low quantum well widths and/or low density. We add here that because of the numerical uncertainty in the thermodynamic energy differences and our simple model for the finite width, the critical value of $1.5\ell$ should be taken only as a first estimate. 

\section{$\mathbb{Z}_n$ state in bilayer graphene}

We next ask if similar physics can appear elsewhere. We expect to find a transition in the zeroth Landau level of bilayer graphene (BLG) as a function of the magnetic field. The zeroth Landau level of BLG is exactly equivalent to the LLL of GaAs when the magnetic field is infinite and continuously interpolates to the SLL of GaAs as the magnetic field is decreased. As such, we expect that a $\mathbb{Z}_n$ state is stabilized below a critical field and the Laughlin state is favored above the critical field.

The Coulomb interaction between electrons can be parameterized using Haldane pseudopotentials $V_m$~\cite{Haldane83}, which is the energy of two electrons in a relative angular momentum state $m$ in the disk geometry. The pseudopotentials in the zeroth LL of bilayer graphene are given by
\begin{equation}
V^{0-{\rm BLG}}_m(\theta)  = \int_{0}^{\infty} dq~F^{0-{\rm BLG}}(\theta,q)e^{-q^2}L_m(q^2).\label{BLGVm}
\end{equation}
where the Fourier transformed form factor is \cite{Apalkov11}
\begin{equation}
F^{0-{\rm BLG}}(\theta,q) = \Bigg[\sin^{2}(\theta)L_{1}\Big(\frac{q^2}{2}\Big)+\cos^{2}(\theta)L_{0}\Big(\frac{q^2}{2}\Big) \Bigg]^2.
\label{eq:ZerothLL_BLG_form_factor}
\end{equation}
Here we have set the magnetic length to unity, $L_r(x)$ is the $r$th order Laguerre polynomial and $\theta$ is a parameter that varies between $0$ and $\pi/2$ to control the relative proportion of the $n=0$ and $n=1$ LLs in the two-component wave function. At $\theta=0$ the form factor is that of the LLL of GaAs while for $\theta = \pi/2$, it is exactly the form factor for the SLL in GaAs. At the mid-way point, $\theta =\pi/4$, the form factor is that of the $n=1$ LL in monolayer graphene~\cite{Balram15c}. The value of $\theta$ is related to the magnetic field by $\tan^2{(\theta)}\propto \ell /(\hbar v_F)\propto 1/\sqrt{B}$, where $v_F$ is the Fermi velocity. For very large magnetic fields, we anticipate the physics of the LLL of GaAs. As the magnetic field is lowered, we first expect to see the physics of monolayer graphene appear (which has been shown to be well described by the composite fermion theory~\cite{Balram15c}) that eventually gives way to states exhibiting the physics of the SLL of GaAs at very small magnetic fields.

\begin{figure}[tbhp]
\includegraphics[width=0.4\textwidth]{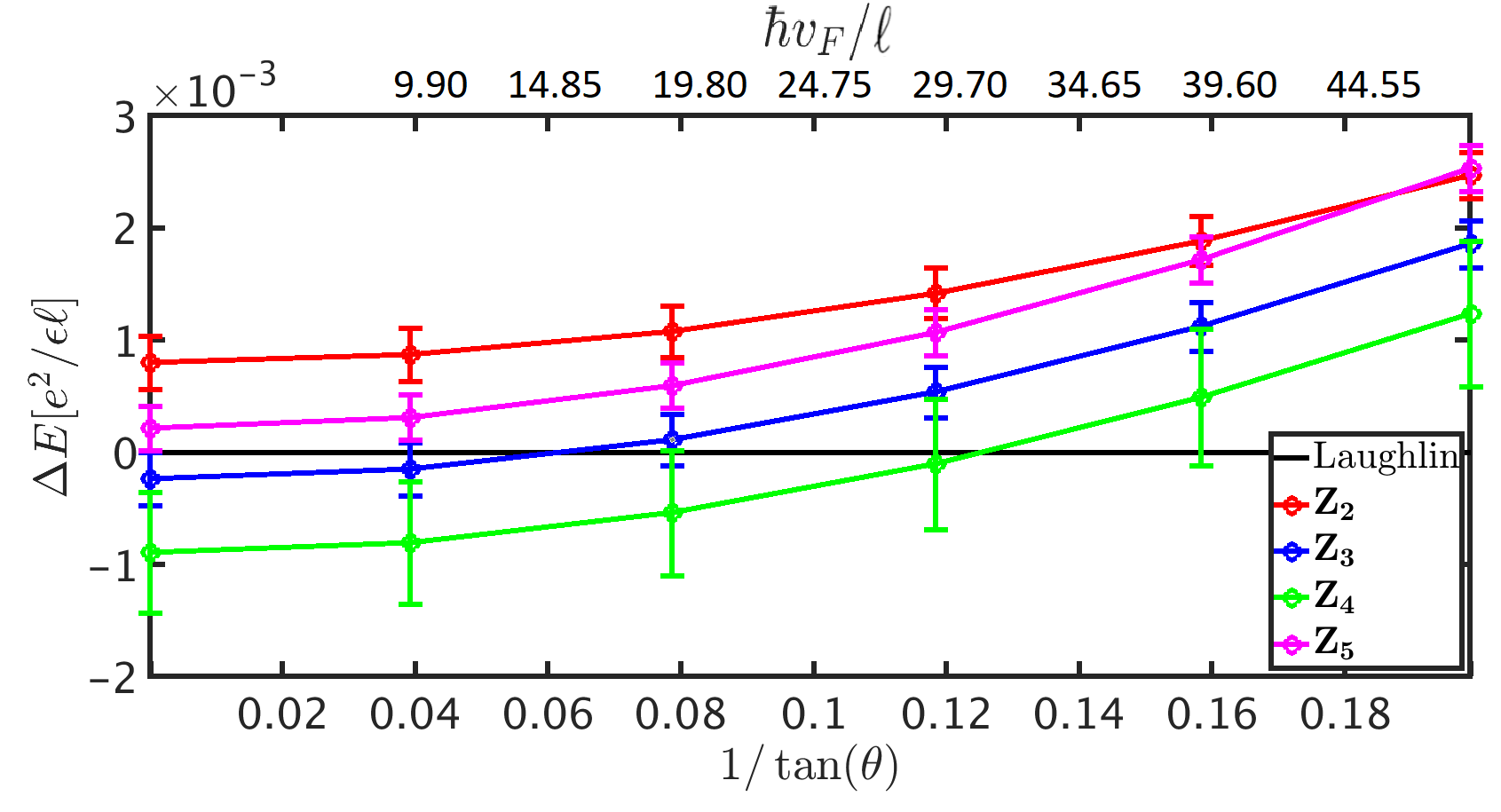}
\caption{Energies of $\mathbb{Z}_n$ states in bilayer graphene, measured relative to the energy of the Laughlin state, as a function of the tangent of the mixing angle $\theta$. All energies represent thermodynamic limits. The $\mathbb{Z}_4$ state is seen to be favored for $\theta \gtrapprox 1.45$. The top axis shows $\hbar v_F/\ell$ in units of meV (see text for relation between $\theta$ and $\hbar v_F/\ell$). Energies are shown only in the vicinity of the transition. }
\label{fig:BiGr}
\end{figure}

We construct an effective interaction as shown in Appendix~\ref{sec: effInt}, and obtain the thermodynamic energies of various candidate states as a function of $\theta$.  The angle $\theta$ is related to measurable quantities through $\tan{\theta}=  t \ell/(\sqrt{2}\hbar v_F)$ where $t$ is the hopping integral and $v_F$ is the Fermi velocity~\cite{Apalkov11}. Taking $t\sim 350\text{meV}$, as obtained from DFT calculations at zero magnetic field~\cite{Jung14}, we obtain $\hbar v_F/\ell = 350\text{meV}/(\sqrt{2}\tan{\theta})$. The top axis in Fig.~\ref{fig:BiGr} shows $\hbar v_F/\ell$ in units of meV.  We find that the transition from the $\mathbb{Z}_{4}$ to the Laughlin state occurs approximately at $\hbar v_F/\ell\sim30$ meV. For graphene, with a typical Fermi velocity of $10^6$ $m/s$, this corresponds to a magnetic field strength of $B\approx 1.4$T.

\section{Landau level mixing and spin}

It is natural to ask if $\mathbb{Z}_n$ parton states can be relevant for $\nu=1/3$ in the LLL. We have performed extensive calculations as a function of quantum well width and density, also including LL mixing. We use a self-consistent LDA calculation to determine the transverse electron density at zero magnetic field at several electron densities and quantum well widths. We further include LL mixing through the so-called fixed phase diffusion Monte Carlo method~\cite{Ortiz93, Zhang16, Zhao18}. For all parameters we have considered, the Laughlin state remains the lowest energy state. The detailed results are given in Appendix~\ref{sec: ThermoLim}.

Recent experiments have mapped out the spin polarization of the SLL in GaAs quantum wells~\cite{Yoo19}. They observe an anomalous spin depolarization between fillings factors 1/5 and 1/3. The Laughlin state is fully spin polarized, but $\mathbb{Z}_n$ states with $n>1$ allow for the possibility of spin-unpolarized or spin-partially polarized states. The generalization is analogous to that for Jain CF states to spin-singlet or partially spin-polarized states~\cite{Wu93, Park98, Jain07, Balram15a, Balram17}. Specifically, the $\mathbb{Z}_n$ state can be generalized to include spin as 
\begin{eqnarray}
\Psi_{1/m}^{\mathbb{Z}_n(n,0;\bar{n}_\uparrow,\bar{n}_\downarrow)} &=& \mathcal{P}_{\rm LLL}\Phi_n\Phi_{\bar{n}_\uparrow,\bar{n}_\downarrow}\Phi_1^m \nonumber \\ 
&\sim & \Psi_{n/(2n+1)} \Phi_1^{m-4} \mathcal{P}_{\rm LLL} \Phi^*_{n_\uparrow,n_\downarrow} \Phi_1^2 
\label{Znspin}
\end{eqnarray}
where $n =n_\uparrow + n_\downarrow$, and $\Phi_{n_\uparrow,n_\downarrow}$ represents the state with $n_\uparrow$ spin-up and $n_\downarrow$ spin-down filled Landau levels. These wave functions can be shown to satisfy the Fock cyclic conditions~\cite{Jain07}.
In the above wave function, we have made the $\bar{n}$-parton spinful. An analogous wave function $\Psi_{1/m}^{\mathbb{Z}_n(n_\uparrow,n_\downarrow;\bar{n},0)}$ can be written where the $n$-parton is endowed with spin. Which configuration is preferred depends on the interaction. Our detailed calculations, shown in Appendix~\ref{sec: ThermoLim}, demonstrate that the fully spin-polarized states have better variational energies for all interactions considered in this article.

\section{Discussion}
\label{sec: discussion}
Our work was motivated by an apparent discrepancy between theory and experiment for the FQHE at $\nu=7/3$: while theory finds the $\mathbb{Z}_4$ parton state to have lower energy than the Laughlin state, experiments are consistent with the latter. We find that when we take into account finite width corrections, there is a transition from the $\mathbb{Z}_4$ state into the Laughlin state at width $\sim 1.5 \ell$. 

All experimental observations of the 7/3 state appear to be for larger widths and thus fall in the region where the Laughlin state is favored. (Large mobilities, necessary for an observation of the 7/3 state, are typically obtained for relatively wide quantum wells because that minimizes the effect of interface roughening. One may alternatively go to low densities. The 7/3 state has been observed at very low densities~\cite{Pan14, Samkharadze17}, but even there, with 7/3 state occurring at $B\approx 0.9$T, the width of 65 nm translates  approximately into 2.5 $\ell$.) It may be possible to decrease both the quantum well width and the density to get into the regime where the $\mathbb{Z}_4$ state is predicted. If a phase transition is observed (for example, by gap closing and reopening as a function of the density), it would provide evidence in favor of an unconventional $\mathbb{Z}_n$ state at small widths, and also of the role of large width in stabilizing the Laughlin state at 7/3. 

We note that a previous exact diagonalization calculation has also shown that the magnetoroton branch, absent at zero width, appears by the time the quantum well width is three magnetic lengths~\cite{Jolicoeur17}. Another exact diagonalization study has shown that finite width stabilizes the 7/3 Laughlin state~\cite{Balram20}.

An additional experimental quantity, namely the chiral central charge, can be measured in thermal Hall conductance measurements and can sometimes distinguish between states with different topological content~\cite{Kane97, Cappelli02}. For the $\mathbb{Z}_n$ states, the chiral central charge is independent of the value of $n$ and so the thermal Hall conductance is predicted to be the same for all of these states. The measured value of the thermal Hall conductance at $7/3$ is consistent with all of these states \cite{Banerjee17b}. 

The Hall viscosity of all $\mathbb{Z}_n$ states is identical because they all have the same shift. These states are, however, not topologically equivalent as they have different topological entanglement entropies~\cite{Balram19a}. The clearest experimental signature distinguishing the states will be the charge of the fundamental quasiparticles. The Laughlin state has charge $-e/3$ quasiparticles while the $\mathbb{Z}_n$ state has charge $-e/(3n)$ quasiparticles. These quasiparticles can, in principle, be detected through scanning electron transistor experiments~\cite{Venkatachalam11}. The situation for shot noise experiments is more subtle. As Balram \emph{et al}. argued~\cite{Balram19a}, the $-e/(3n)$ quasiparticles are gapped at the edge and only the $-e/3$ quasiparticles can be excited at arbitrarily low temperatures.  It may be possible, however, that the $-e/(3n)$ quasiparticles become relevant in shot noise experiments at somewhat elevated temperatures (or voltage bias).

We note that the so-called anti-Read-Rezayi $4$-cluster (aRR$4$) state~\cite{Read99} also provides a plausible candidate wave function for the $7/3$ FQHE~\cite{Peterson15}. The energy of the aRR$4$ state is equal to the energy of the Laughlin state within numerical uncertainty~\cite{Peterson15}, in contrast to our $\mathbb{Z}_4$ state which has lower energy than Laughlin's. Furthermore, the aRR$4$ state has overlaps of 0.77 and 0.59 with the exact ground state for 10~\cite{Kusmierz18} and 12 particles, whereas the $\mathbb{Z}_2$ state has an overlap of 0.87 for 10 particles and $\mathbb{Z}_3$ has an overlap of 0.93 for 9 particles~\cite{Balram19a}. (The $\mathbb{Z}_4$ state requires a minimum of 16 particles, for which we cannot obtain overlaps.) Finally, assuming equilibration of all edge modes, the thermal Hall measurements at 7/3 are inconsistent with the chiral central charge of the aRR$4$ state~\cite{Banerjee17b}. 

\begin{acknowledgments}
The work at Penn State was supported by the U. S. Department of Energy, Office of Basic Energy Sciences, under Grant no. DE-SC0005042. Some portions of this research were conducted with Advanced CyberInfrastructure computational resources provided by The Institute for CyberScience at The Pennsylvania State University. Some of the numerical calculations reported in this work were carried out on the Nandadevi supercomputer, which is maintained and supported by the Institute of Mathematical Science’s High Performance Computing center. One of us (Th.J.) acknowledges CEA-DRF for providing CPU time on the supercomputer COBALT at GENCI-CCRT.
\end{acknowledgments}

\bibliography{../biblio_fqhe}
\bibliographystyle{apsrev}

\appendix

\section{Effective Interactions}
\label{sec: effInt}
The system of electrons in any given Landau level (LL) interacting with the Coulomb interaction is formally equivalent to the system of electrons in the lowest LL (LLL) interacting with an effective interaction that has the same pseudopotentials in the LLL as the Coulomb interaction does in the given LL~\cite{Haldane83}. This allows us to work within the LLL, which is convenient because various trial wave functions are most readily constructed within the LLL. Such effective interactions have been constructed for zero as well as finite width systems~\cite{Park99b, Shi08, Toke08}. Below we present our construction of the effective interactions that capture the effect of finite width in the LLL and second LL (SLL) of conventional semiconductors such as GaAs. We also construct effective interactions that describe the physics of the zeroth LL of bilayer graphene (BLG) as a function of the magnetic field.

\subsection{Finite width in LLs of GaAs}
In the LLL, we use a self-consistent LDA calculation to determine the transverse electron density at zero magnetic field at several electron densities and quantum well widths \cite{Park99b}. We consider densities from $1\times 10^{10}~\text{cm}^{-2}$ to $30\times 10^{10}~\text{cm}^{-2}$ and quantum well widths from 18 nm to 70 nm. In the SLL, we approximate the confining potential as an infinite square well that results in a sinusoidal transverse wave function. The effective interaction for this system can be found in Ref.~\cite{Toke08}.

\subsection{Bilayer Graphene}
In Sec. IV of the main text we introduced the pseudopotentials $\{V^{0-{\rm BLG}}_m(\theta)\}$ in the zeroth LL of bilayer graphene. The interaction given in Eq.~(3) of the main text can be integrated analytically to obtain
\begin{eqnarray}
\label{Vm0BLG}
V^{0-{\rm BLG}}_m(\theta) &=& \frac{\sqrt{\pi}}{32}\Bigg [16~_2F_1 \left(\frac{1}{2},-m;1;1 \right) \\
&&- 8~_2F_1 \left(\frac{3}{2},-m;1;1 \right)\sin^2(\theta) \nonumber \\ 
&& + 3~_2F_1 \left(\frac{5}{2},-m;1;1 \right)\sin^4(\theta)\Bigg ] \nonumber
\end{eqnarray}
where $_2F_1$ is the Gauss hypergeometric function. We propose the following real space effective interaction to describe the physics of the zeroth LL of bilayer graphene in the LLL 
\begin{equation}
V_{\rm eff}(r,\theta) = \frac{1}{r} + \frac{\sin^2(\theta)}{\sqrt{r^6+1}} + \frac{2.25\sin^4(\theta)}{\sqrt{r^{10}+10}} + \sum_{i=0}^{M} C_i r^{2i}e^{-r^2}, 
\label{Veff}
\end{equation}
where $r$ is in units of $\ell$ and the $C_{i}$'s are a set of $M+1$ fitting parameters. The coefficients of the first three terms, namely $1$, $\sin^2(\theta)$ and $2.25\sin^4(\theta)$ are fixed by the long-range part of the Coulomb interaction. We use the same functional form given in Eq.~(\ref{Veff}) for both fully polarized (or ``spinless") electrons, for which only the odd $m$ pseudopotentials are relevant, and non-fully polarized (or ``spinful") electrons, for which all the pseudopotentials are relevant. For the spinful case, we fit the first 7 pseudopotentials ($m=0,\cdots,6$) of Eq.~(\ref{Vm0BLG}) with that of the effective interaction given in Eq.~(\ref{Veff}) to determine the coefficients $C_{0}, C_{1}, \cdots, C_{6}$. This interaction can also be used for spinless states, but for these states we find it more convenient to
keep only three fitting parameters $C_{0}, C_{1}$ and $C_{2}$ in the effective interaction and determine them by fitting the first 3 relevant pseudopotentials $(m=1,3,5)$. The reason is that the effective interaction generated for the spinless case using this procedure is already very accurate; fitting a larger number of pseudopotentials leads to a highly oscillating effective interaction, thus accentuating numerical errors in the energy calculations. For the fully polarized states of our interest, we have calculated energies using both interactions, and confirmed that they are consistent (within error bars). 

\begin{figure*}[htpb]
\includegraphics[width = 0.45\textwidth]{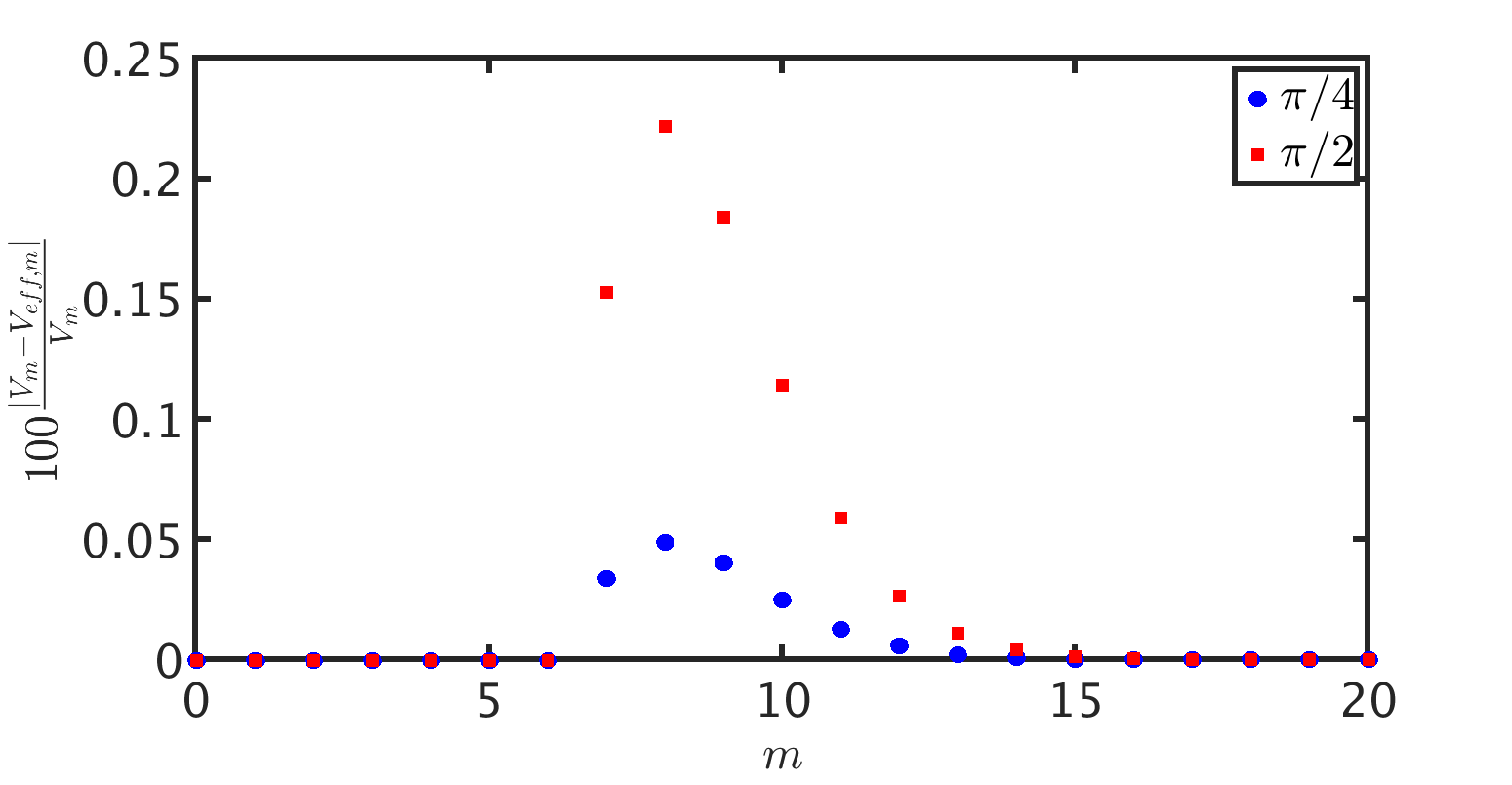}
\includegraphics[width = 0.45\textwidth]{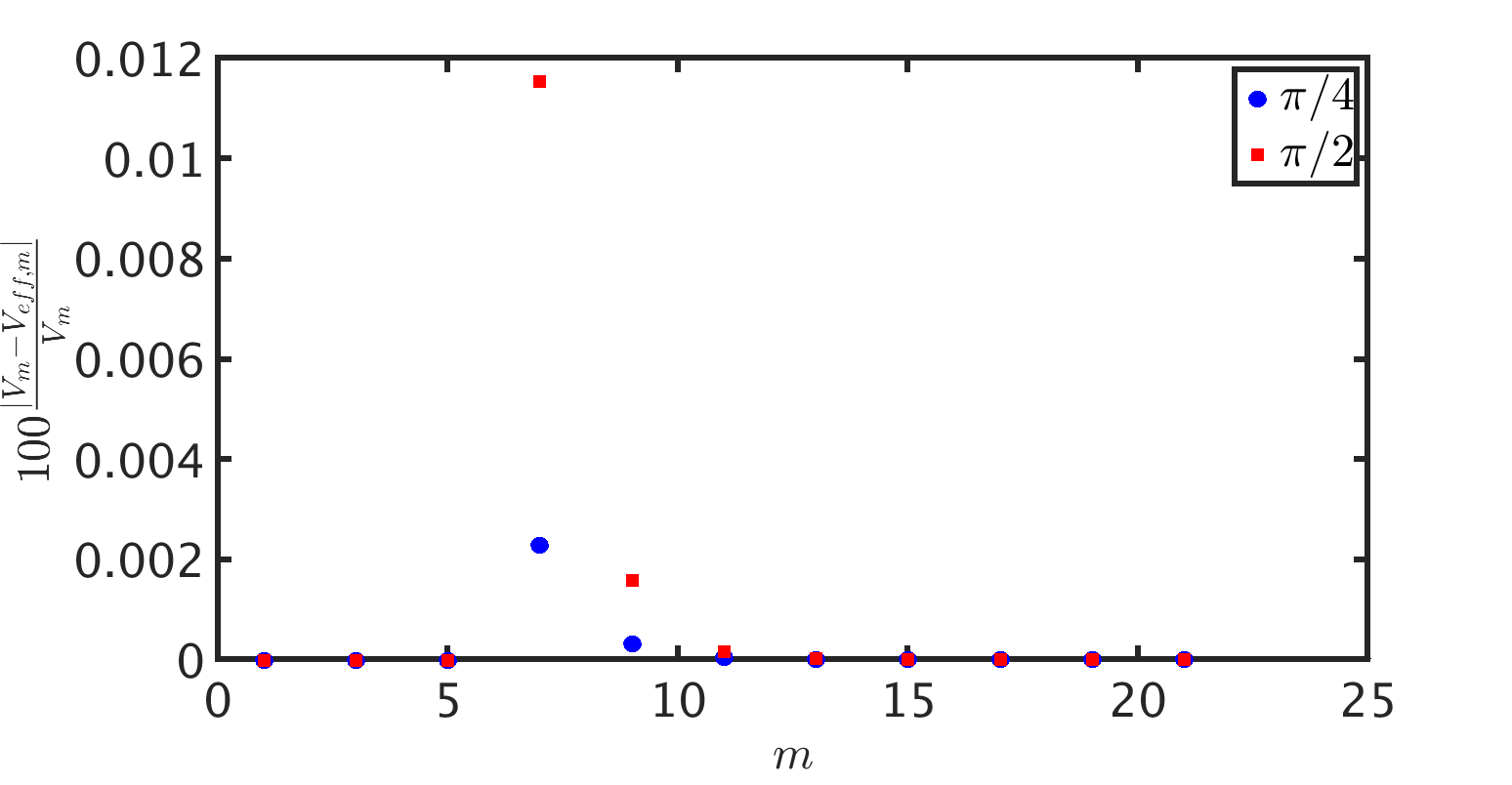}
\caption{Deviation of the pseudopotentials of the effective interactions from true pseudopotentials in the zeroth Landau level of bilayer graphene. Panels (a) and (b) correspond to the interactions for spinful and spinless electrons, respectively; for the former both odd and even pseudopotentials are relevant, whereas for the latter only odd pseudopotentials are relevant. We present data for $\theta = \pi/4$ (blue circles) and $\pi/2$ (red squares). The error for intermediate values of $\theta$ smoothly interpolates between these two extreme values of our interest. The $\theta = \pi/4$ and $\theta = \pi/2$ interactions correspond to the interaction in the $n=1$ LL of monolayer graphene and the SLL of GaAs, respectively.}
\label{fig:err}
\end{figure*}

For completeness, in Tables~\ref{tab: CoeffSp} and \ref{tab: Coeff} we have tabulated the coefficients $\{C_{i}\}$ for the spinful and spinless effective interactions. As seen in Fig \ref{fig:err}, the deviation of the pseudopotentials of the effective interaction from the true pseudopotentials [given in Eq.~(\ref{Vm0BLG})] is always below 0.3$\%$, confirming that our effective real space interaction accurately captures the physics in the zeroth LL of bilayer graphene for all values of $\theta$ (The value of $\theta$ can be tuned by varying the magnetic field.). Shi~\emph{et al.} ~\cite{Shi08} used the same procedure to construct an effective interaction to simulate the physics of the fully spin-polarized SLL states in the LLL; our interaction for spinless systems matches theirs for $\theta = \pi/2$. An effective interaction for spinful electrons in the $n=1$ LL of monolayer graphene was earlier constructed in a similar fashion in Ref.~\cite{Balram15c}; our interaction agrees with that of Ref.~\cite{Balram15c} for $\theta = \pi/4$.

\begin{table*}[]
\begin{centering}
\begin{tabular}{|l|l|l|l|l|l|l|l|}
\hline
$\theta$ &  $C_0$ & $C_1$ & $C_2$ & $C_3$ & $C_4$ & $C_5$ & $C_6$\\
\hline
$\pi/4$ & 42.552118895 & -284.79594413 & 376.9515623 & -179.3196865 & 36.582878699 & -3.23227919 & 0.100424064 \\ \hline
$21\pi/80$ & 52.355216896 & -344.41619237 & 453.2479057 & -214.9925686 & 43.784369086 & -3.86408630 & 0.119954345 \\ \hline
$22\pi/80$ & 63.029793451 & -409.04246282 & 535.8091364 & -253.5598803 & 51.565847438 & -4.54651997 & 0.141043830 \\ \hline
$23\pi/80$ & 74.484207658 & -478.12738090 & 623.9395653 & -294.6975250 & 59.862033129 & -5.27386166 & 0.163515945 \\ \hline
$24\pi/80$ & 86.605498835 & -551.00087294 & 716.7898308 & -338.0103786 & 68.593409224 & -6.03914995 & 0.187155830 \\ \hline
$25\pi/80$ & 99.261125521 & -626.88030901 & 813.3696705 & -383.0382094 & 77.667411251 & -6.83428442 & 0.211713546 \\ \hline
$26\pi/80$ & 112.30120610 & -704.88353603 & 912.5643142 & -429.2632732 & 86.979951610 & -7.65015890 & 0.236908181 \\ \hline
$27\pi/80$ & 125.56121220 & -784.04451876 & 1013.154144 & -476.1194174 & 96.417246762 & -8.47682112 & 0.262432767 \\ \hline
$28\pi/80$ & 138.86505467 & -863.33124095 & 1113.837188 & -523.0024955 & 105.85790672 & -9.30365525 & 0.287959912 \\ \hline
$29\pi/80$ & 152.02849210 & -941.66546226 & 1213.253932 & -569.2818539 & 115.17523982 & -10.1195832 & 0.313148002 \\ \hline
$30\pi/80$ & 164.86278371 & -1017.9438792 & 1310.013895 & -614.3126321 & 124.23972015 & -10.9132805 & 0.337647844 \\ \hline
$31\pi/80$ & 177.17850208 & -1091.0602024 & 1402.723346 & -657.4485910 & 132.92156104 & -11.6734005 & 0.361109591 \\ \hline
$32\pi/80$ & 188.78941695 & -1159.9276378 & 1490.013529 & -698.0551737 & 141.09333491 & -12.3888038 & 0.383189799 \\ \hline
$33\pi/80$ & 199.51635936 & -1223.5012479 & 1570.568726 & -735.5224941 & 148.63257863 & -13.0487854 & 0.403558436 \\ \hline
$34\pi/80$ & 209.19097536 & -1280.7996689 & 1643.153519 & -769.2779501 & 155.42432337 & -13.6432961 & 0.421905691 \\ \hline
$35\pi/80$ & 217.65928092 & -1330.9256744 & 1706.638594 & -798.7981650 & 161.36348985 & -14.1631513 & 0.437948421 \\ \hline
$36\pi/80$ & 224.78493416 & -1373.0851016 & 1760.024496 & -823.6199775 & 166.35709251 & -14.6002238 & 0.451436086 \\ \hline
$37\pi/80$ & 230.45214761 & -1406.6036925 & 1802.462760 & -843.3502199 & 170.32620076 & -14.9476150 & 0.462156023 \\ \hline
$38\pi/80$ & 234.56817133 & -1430.9414532 & 1833.273930 & -857.6740553 & 173.20761110 & -15.1998010 & 0.469937948 \\ \hline
$39\pi/80$ & 237.06528826 & -1445.7041910 & 1851.962037 & -866.3616757 & 174.95519048 & -15.3527499 & 0.474657572 \\ \hline
$\pi/2$ &  237.90227427 & -1450.6519568 & 1858.225190 & -869.2732028 & 175.54085927 & -15.4040074 & 0.476239246 \\ \hline
\end{tabular}
\end{centering}
\caption{Coefficients of the effective interaction given in Eq.~(\ref{Veff}) for bilayer graphene as a function of the angle $\theta$ for a system of spinful electrons.}
\label{tab: CoeffSp}
\end{table*}

\begin{table}[]
\begin{tabular}{|l|l|l|l|}
\hline
$\theta$ & $C_0$ & $C_1$ & $C_2$\\
\hline
$\pi/4$ & -3.500190774   & 2.2905277281  & -0.2760897478 \\ \hline
$21\pi/80$ & -4.3645703465  & 2.8256689456  & -0.3392114604 \\ \hline
$22\pi/80$ & -5.3086398458  & 3.4087491892  & -0.4079208489 \\ \hline
$23\pi/80$ & -6.3242224208  & 4.0347534781  & -0.4816286458 \\ \hline
$24\pi/80$ & -7.4011939407  & 4.6974943436  & -0.5596088683 \\ \hline
$25\pi/80$ & -8.5276405265  & 5.389707299   & -0.641009981  \\ \hline
$26\pi/80$ & -9.6900618674  & 6.1031738918  & -0.724869275  \\ \hline
$27\pi/80$ & -10.8736158706 & 6.8288696533  & -0.8101301508 \\ \hline
$28\pi/80$ & -12.0623991601 & 7.5571336381  & -0.8956619189 \\ \hline
$29\pi/80$ & -13.2397570448 & 8.2778557024  & -0.9802816689 \\ \hline
$30\pi/80$ & -14.3886158229 & 8.9806772239  & -1.0627777046 \\ \hline
$31\pi/80$ & -15.4918297228 & 9.6552006158  & -1.1419340046 \\ \hline
$32\pi/80$ & -16.5325343885 & 10.2912027588 & -1.2165551382 \\ \hline
$33\pi/80$ & -17.4944986309 & 10.8788473569 & -1.285491054  \\ \hline
$34\pi/80$ & -18.3624661736 & 11.4088912318 & -1.3476611604 \\ \hline
$35\pi/80$ & -19.1224793364 & 11.8728796968 & -1.4020771302 \\ \hline
$36\pi/80$ & -19.7621770063 & 12.263326399  & -1.4478638919 \\ \hline
$37\pi/80$ & -20.2710598417 & 12.5738733751 & -1.4842783109 \\ \hline
$38\pi/80$ & -20.6407164192 & 12.7994275309 & -1.5107251187 \\ \hline
$39\pi/80$ & -20.8650049565 & 12.936270306  & -1.5267697123 \\ \hline
$\pi/2$ & -20.9401862915 & 12.9821379218 & -1.5321475202 \\ \hline
\end{tabular}
\caption{Coefficients of the effective interaction given in Eq.~(\ref{Veff}) for bilayer graphene as a function of the angle $\theta$ for a system of fully polarized electrons.}
\label{tab: Coeff}
\end{table}

\section{Variational Monte Carlo}
\label{sec: VMC}

We evaluate the energy of each trial wave function for our effective interactions using variational Monte Carlo (VMC). VMC allows us to consider many different interactions during the same sampling procedure and so we can quickly determine the energy for the LLL, the SLL, finite width effective interactions, and bilayer graphene effective interactions.  In each VMC run, we perform 20 million steps and typically estimate the error in the per-particle ground-state energy of each system to be between $10^{-5}$ and $10^{-4}$, except for highly oscillatory interactions for spinful electrons for which the error is $\sim2\times10^{-3}$. The reported energies are the total energies per particle, i.e., include the effective electron-electron interaction, the electron-background interaction, and the background-background interaction. (For the latter two, we use the Coulomb interaction, rather than the effective interaction. However, this is sufficient, because we are only interested in the energy differences.)

There is a systematic correction to the energy as a function of $N$ due to the dependence of the density on $N$; we correct for this by multiplying the energy by $\sqrt{2Q\nu/N}$ \cite{Morf86}. This correction also reduces finite-size corrections that arise from the fact that the non-fully polarized states can occur at different values of the magnetic monopole strength $2Q$ than the fully polarized states.

Rather than obtaining the thermodynamic limits of the energies, as was done in Ref.~\cite{Balram19a}, we find it more convenient to directly obtain the thermodynamic limits for the energy differences. We find that these behave more linearly with $1/N$ than the individual energies.

\section{Diffusion Monte Carlo: Landau level mixing}
\label{sec: DMC}
The electrons resides in the LLL only in the limit of an infinite magnetic field. The Landau level mixing (LLM) parameter $\kappa=\hbar \omega_c/({e^2/\epsilon \ell})$, which is the ratio of the cyclotron energy to the Coulomb energy, is usually of the order of $1$, implying that LLM might have a non-negligible effect on the energy ordering of various states. Several earlier papers have shown that LLM can affect the nature of the ground state~\cite{Zhao18, Zhang16}. One may wonder whether LLM can trigger a phase transition in our system.

The method we use to study the effect of LLM is called the fixed-phase diffusion Monte Carlo (DMC) method. The standard DMC is a Monte Carlo method designed to obtain the ground state of a many-body Schr\"{o}dinger equation by a stochastic method provided that the ground state wave function is real and non-negative. By setting time to an imaginary variable $t \to t=-i \tau$, the Schr\"{o}dinger equation takes the form:
\begin{equation}
-\hbar \partial_\tau \Psi\left(\vec{R}, \tau\right)=(H-E_T)\Psi\left(\vec{R}, \tau\right)
\end{equation}
where $\vec{R}=(\vec{r}_1, \vec{r}_2, ..., \vec{r}_N)$ is the collective coordinate of the system and $E_T$ is a constant energy offset. When $\Psi$ is real and non-negative, this equation can be interpreted as a diffusion equation, with $\Psi$ interpreted as the density distribution of a collection of randomly diffusing walkers. The energy offset $E_T$ is introduced for technical convenience. Briefly, the simulation starts with a trial wave function $\Psi_T$ and performs the diffusion process to obtain a stable state. As the system evolves, all the high-energy components of the state decay exponentially, so the final state only contains the lowest-energy eigenstate, no matter what initial state is chosen (provided that the initial state has a nonzero overlap with the exact ground state). Many excellent articles introducing the details of DMC are available in the literature~\cite{Foulkes01, Mitas98}.

The original DMC cannot be implemented here because the ground state wave function of an FQHE state is a complex function. To overcome this difficulty, the fixed-phase DMC method has been developed\cite{Ortiz93, Zhao18, Zhang16}. In this method, the ground state wave function is written as
\begin{equation}
\Psi( \vec{R})=\left|\Psi( \vec{R})\right|\exp\left[i \phi( \vec{R})\right]
\end{equation}
The phase $\phi( \vec{R})$ is then fixed to be the phase of a known trial wave function $\Psi_T$ and the standard DMC is implemented to solve for the amplitude part $\left|\Psi( \vec{R})\right|$ to obtain the lowest energy state in the chosen phase sector. This amounts to solving the Schr\"{o}dinger equation
\begin{equation}
\begin{aligned}
&H_\text{DMC} \left| \Psi\left(\vec{R}, \tau\right)\right|=\\
& \left( -\sum_{i=1}^N \frac{\hbar^2\nabla_i^2}{2m_e} +V_{\text{DMC}}( \vec{R})-E_T\right) \left|\Psi\left(\vec{R}, \tau\right)\right|=E \left|\Psi\left(\vec{R}, \tau\right)\right|\\
\end{aligned}
\label{eqn_amp}
\end{equation}
with
\begin{equation}
V_{\text{DMC}}( \vec{R})=V( \vec{R})+\frac{1}{2 m_e}\sum_{i=1}^N
\left[ \hbar \nabla_i \phi ( \vec{R})+\frac{e}{c} {\mathbf A}\left({\mathbf r }_i\right) \right]^2 \label{Vdmc}.
\end{equation}
where $V( \vec{R})$ is the interaction energy and $m_e$ is the electron band mass.

The accuracy of the fixed-phase DMC depends on the choice of the phase $\phi( \vec{R})$. In this work, we fix  $\phi( \vec{R})$ by choosing the trial wave functions to be the $\mathbb{Z}_n$ state. It is worth noting that although the trial wave functions themselves are in the LLL, the final states generated by the DMC algorithm are not restricted to be within the LLL, and the LLM parameter $\kappa$ is embedded in this algorithm. 

\section{$\nu=1/3$ in the LLL}
\label{sec: ThermoLim}

\begin{figure*}[htpb]
\includegraphics[width=0.45\textwidth]{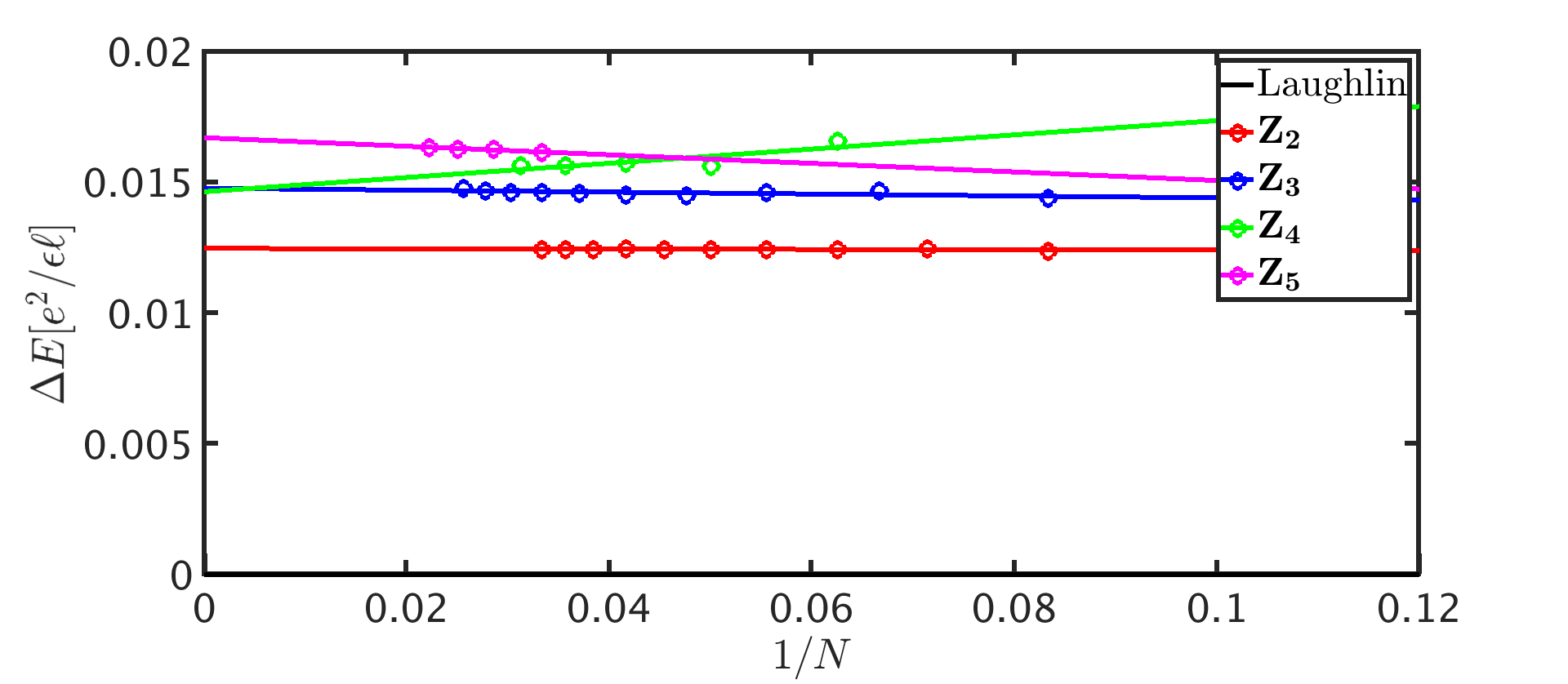}
\includegraphics[width=0.45\textwidth]{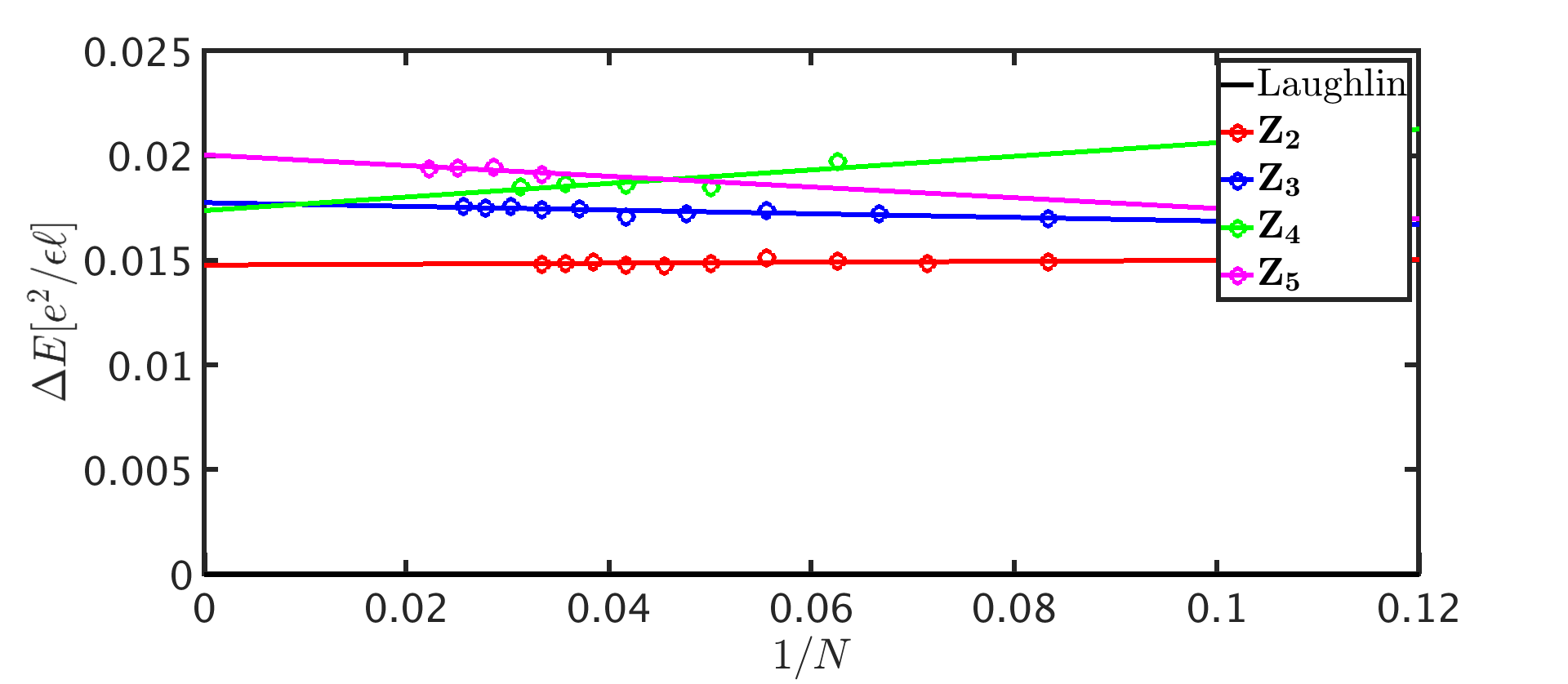}
\caption{Thermodynamic extrapolations of the energy difference between candidate state and the Laughlin state in the LLL (left), and $n=1$ LL of monolayer graphene (right). For both cases, the Laughlin state is the energetically preferred state. }
\label{fig: TLs2}
\end{figure*}

For completeness, we present thermodynamic extrapolations of several $\mathbb{Z}_n$ states at $\nu=1/3$. In all cases, we extrapolate the energy difference between the $\mathbb{Z}_n$ state and the Laughlin state. Fig \ref{fig: TLs2} shows that the thermodynamic extrapolations of the differences for the LLL of GaAs and the $n=1$ LL of monolayer graphene are well fitted by a linear $1/N$ dependence. The Laughlin state is seen to be the lowest energy state. (Results are not shown for the $n=0$ LL of monolayer graphene because it is equivalent to the LLL of GaAs for a zero width system in the absence of LLM.) 

\begin{figure*}[htpb]
\includegraphics[width=0.4\textwidth]{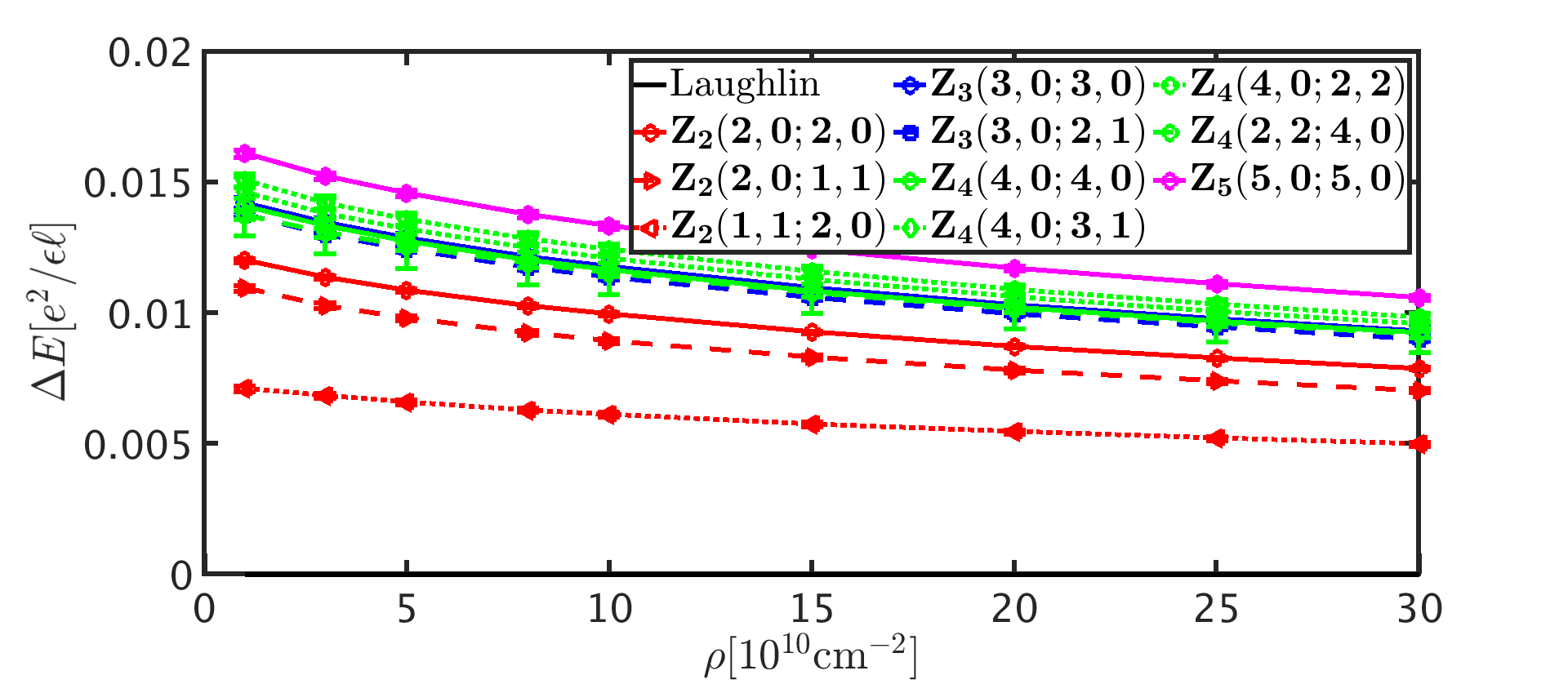}
\includegraphics[width=0.4\textwidth]{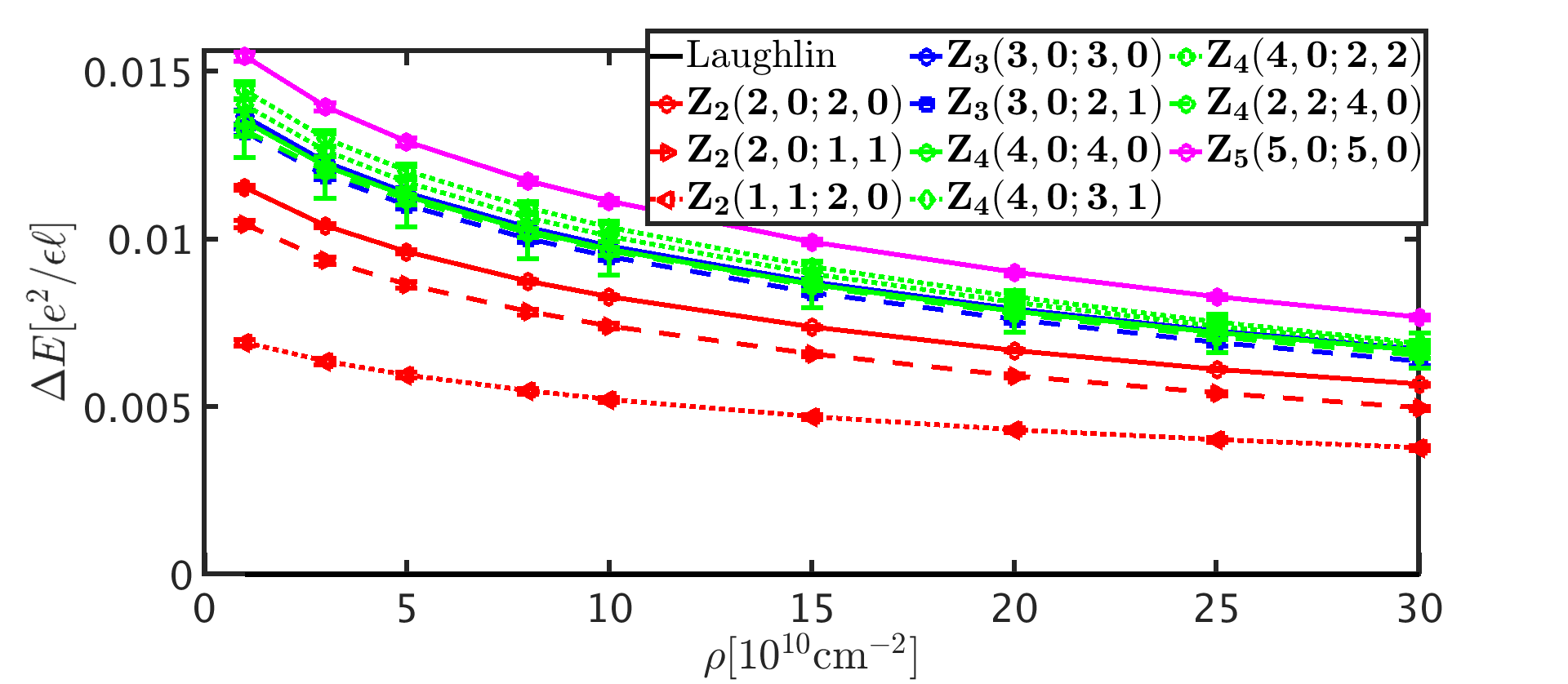}\\
\includegraphics[width=0.4\textwidth]{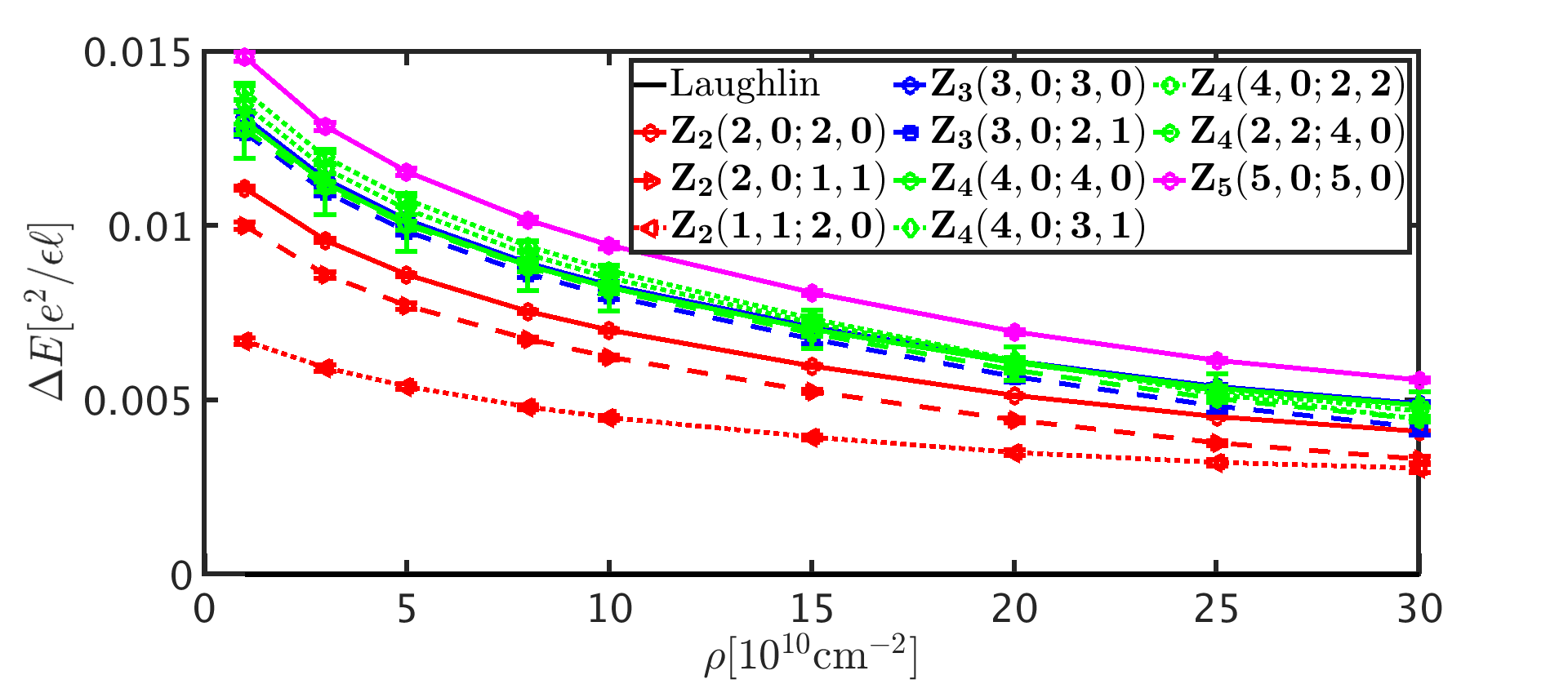}
\includegraphics[width=0.4\textwidth]{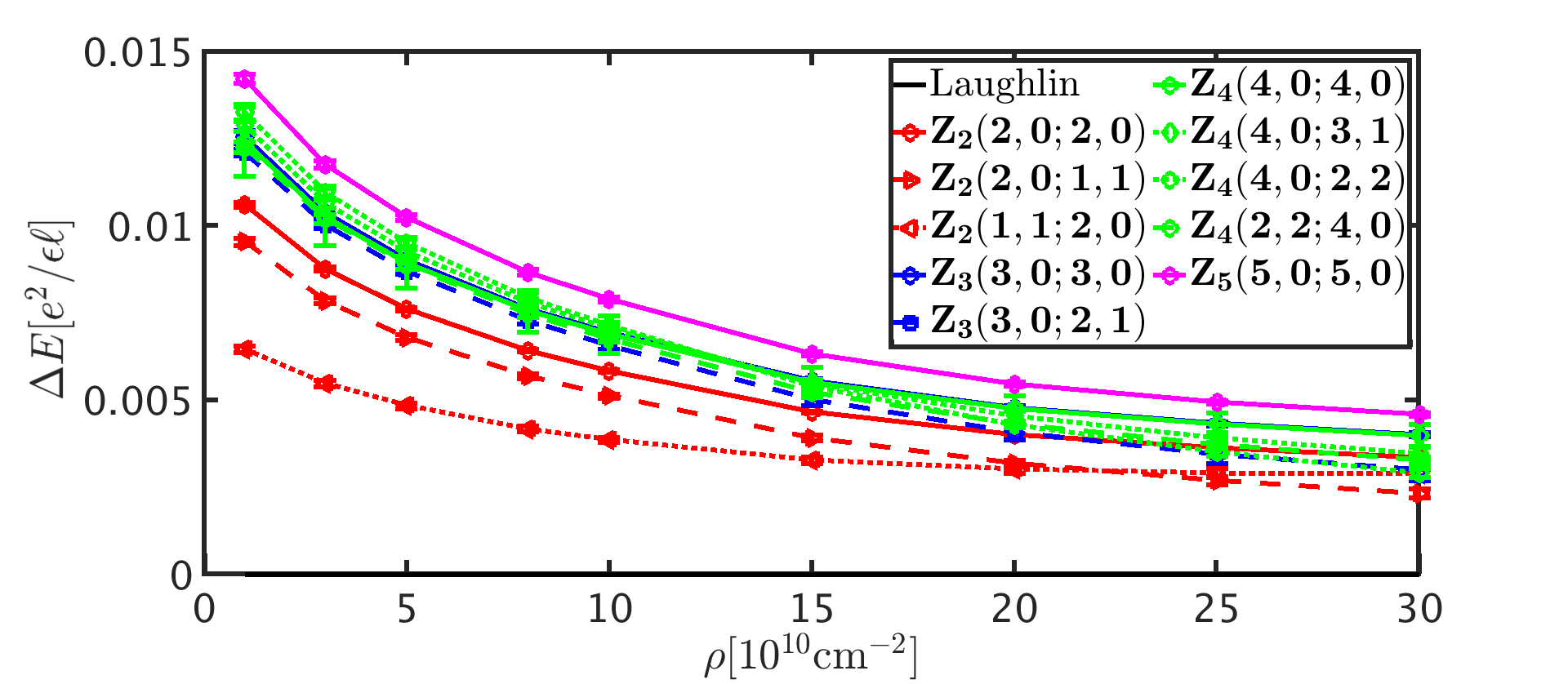}\\
\includegraphics[width=0.4\textwidth]{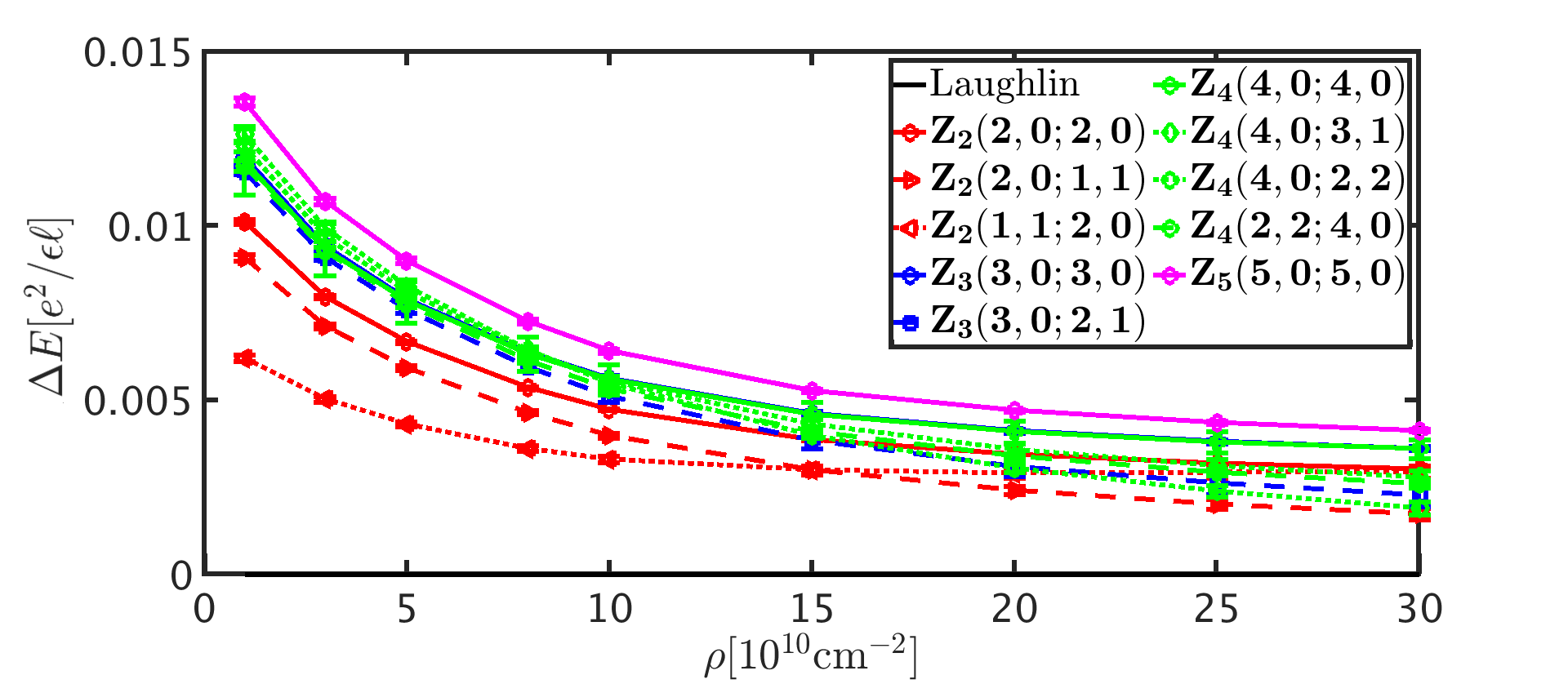}
\includegraphics[width=0.4\textwidth]{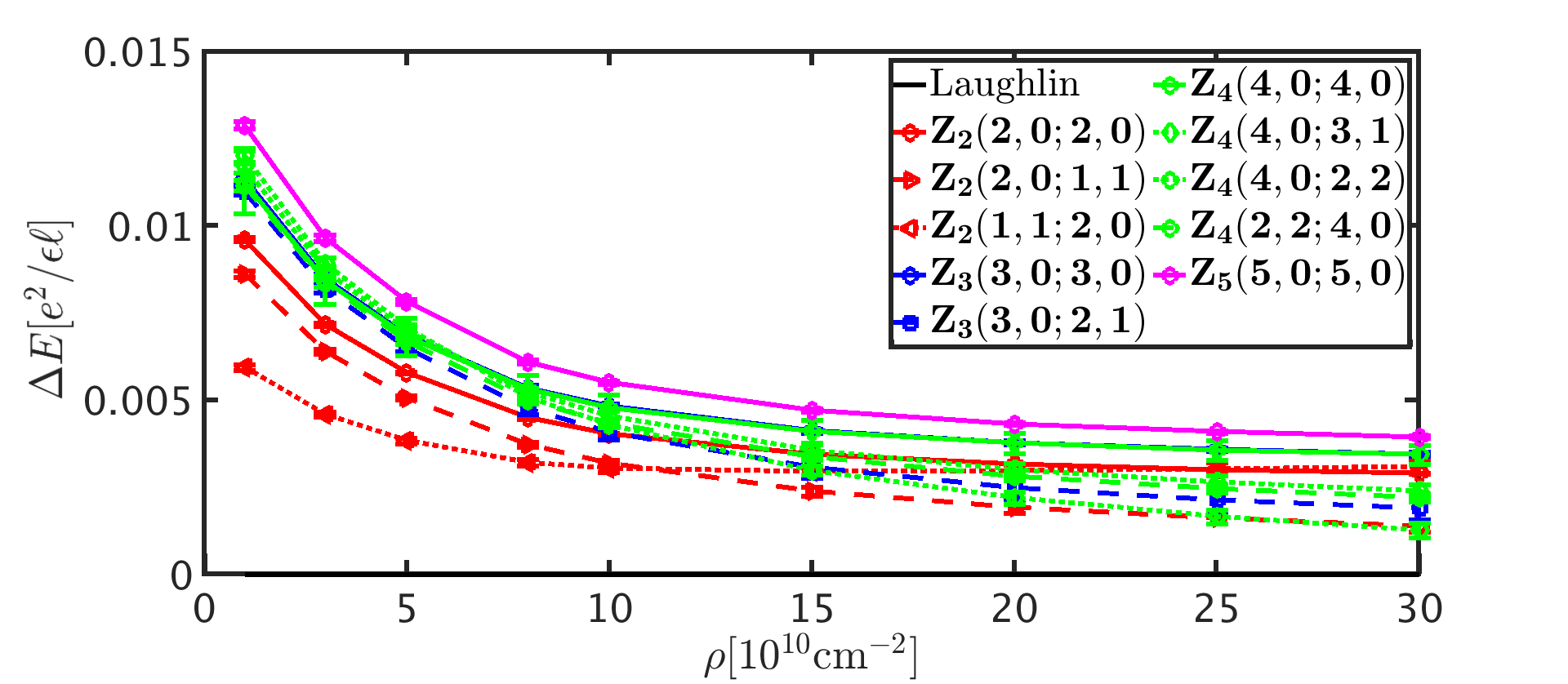}
\caption{Energies of various $\mathbb{Z}_n$ states at $\nu=1/3$ as a function of the quantum well width and the density. $\Delta E$ is the energy difference between the $\mathbb{Z}_n$ state and the Laughlin state, and $\rho$ is the areal electron density in units of $10^{10}$cm$^{-2}$. Panels (a)-(f) show results for quantum wells of width equal to 18 nm, 30 nm, 40 nm, 50 nm, 60 nm, and 70 nm, respectively. The Laughlin state remains the ground state for all parameters considered here.}
\label{fig: FWLLL}
\end{figure*}

Fig.~\ref{fig: FWLLL} presents energies of fully and partially spin polarized $\mathbb{Z}_n$ states in the LLL, considering finite width effects (but no LLM). We find that the energy differences decrease with increasing width and density, as expected. However, the Laughlin state remains the ground state, and there is no transition as a function of either the density or quantum well width for the parameters that we have studied.

\begin{figure*}[htpb]
\includegraphics[width=0.49\textwidth]{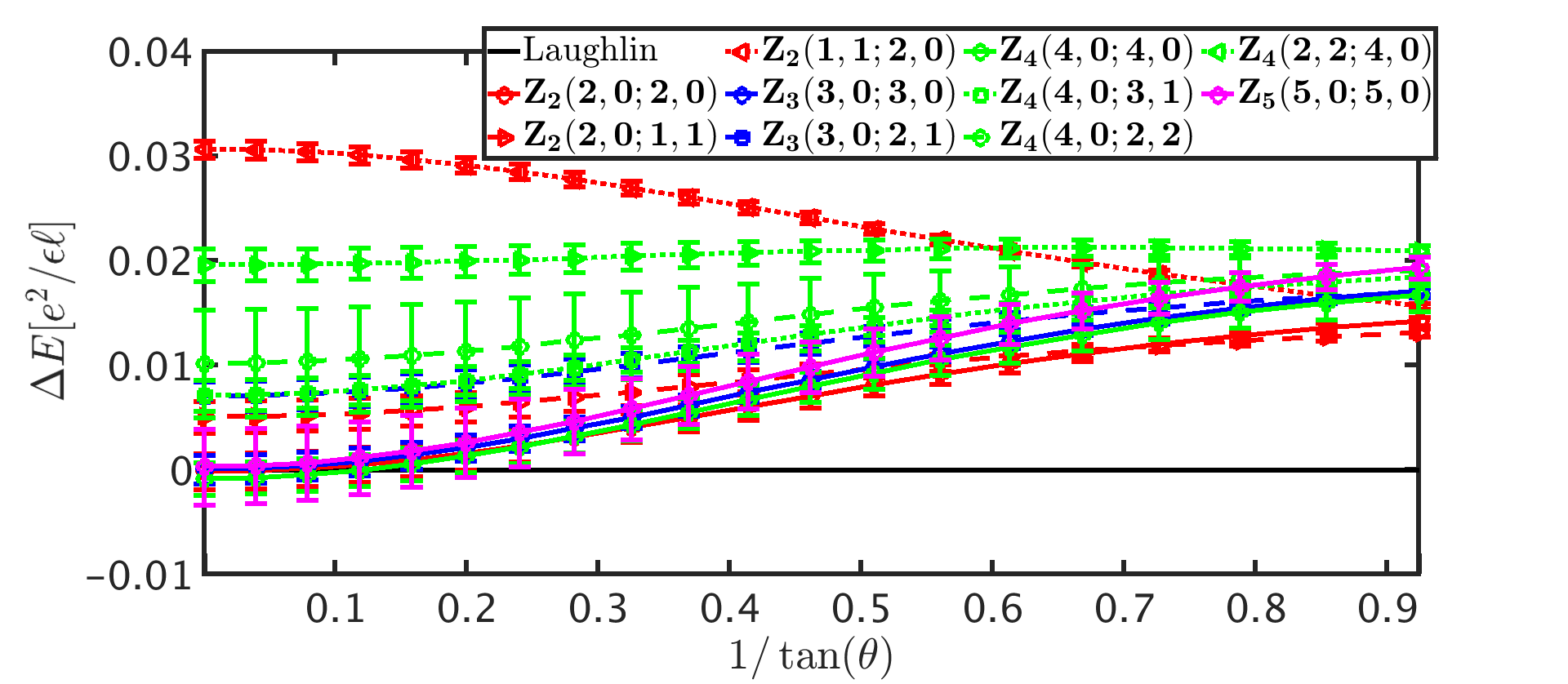}
\includegraphics[width=0.49\textwidth]{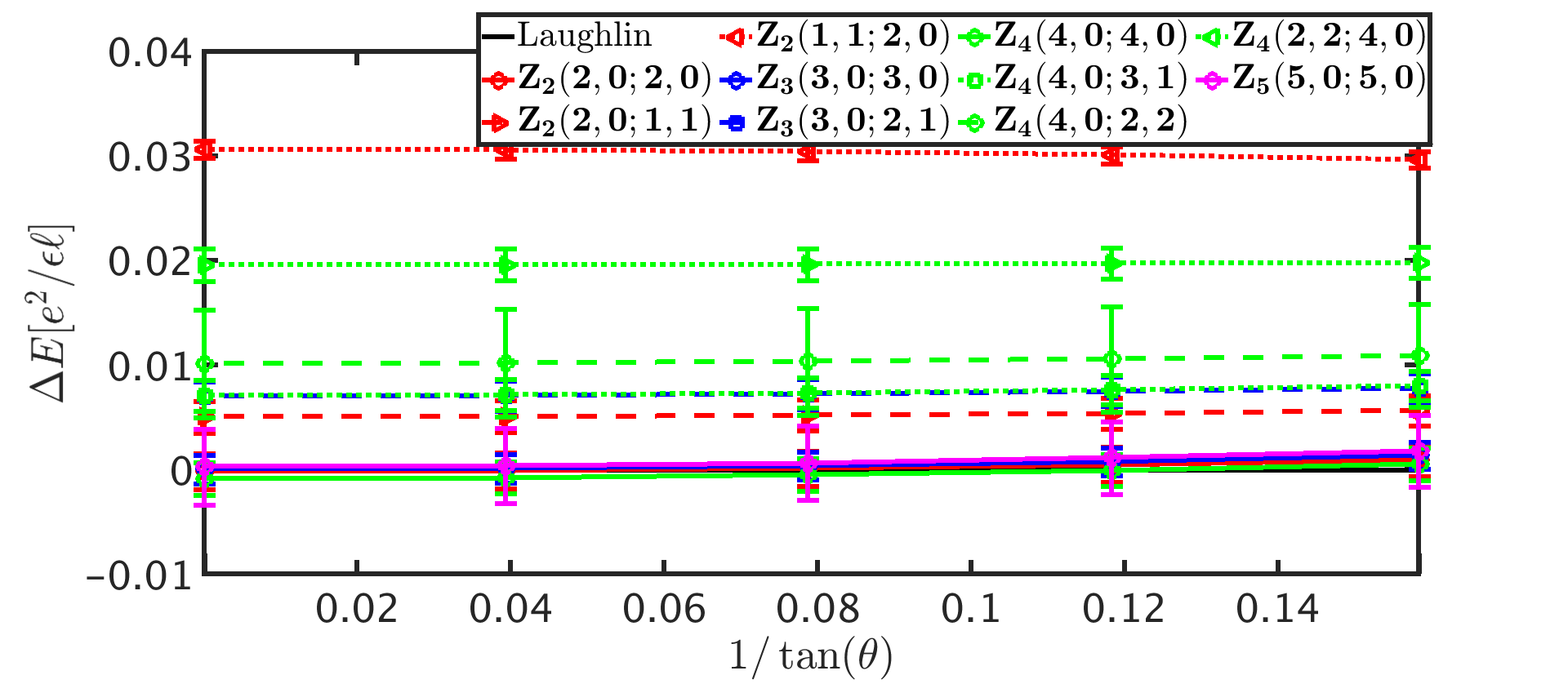}
\caption{Energies of $\mathbb{Z}_n$ states, including spin-unpolarized wave functions, for bilayer graphene as a function of mixing angle $\theta$. All energies are quoted relative to the energy of the Laughlin state. The statistical error in the Monte-Carlo energies of the finite systems grows as $\theta$ is increased. Even with this relatively high error, it is clear that none of the spin unpolarized wave functions is energetically relevant, even in the limit of vanishing Zeeman energy. }
\label{fig: BiGrSp}
\end{figure*}

We have also studied the energy differences between the spinful $\mathbb{Z}_n$ states as a function of the magnetic field in bilayer graphene (assuming no LLM). As seen in Fig~\ref{fig: BiGrSp}, the spin-singlet or partially spin-polarized states are not relevant in the whole parameter regime available.

We next investigate whether Landau level mixing can stabilize a non-Laughlin state at $\nu=1/3$ in the LLL. We report here on our results of fixed-phase DMC calculation for several $\mathbb{Z}_n$ states at $\nu=1/3$. (We are not able to study LLM for states at $\nu=7/3$ because DMC requires keeping all electrons -- not just electrons in the second LL -- which makes the computation of the thermodynamic energies prohibitively demanding.)
	
We study the most prominent $\mathbb{Z}_n$ states in the LLL, which, using the notation defined in Eq.~(5) of the main text, are $\mathbb{Z}_2\left(2,0;2,0\right)$, $\mathbb{Z}_2\left(1,1;2,0\right)$, $\mathbb{Z}_2\left(2,0;1,1\right)$ and $\mathbb{Z}_3\left(3,0;3,0\right)$. We only consider zero well-width. A technical point is worth mentioning. To construct the above wave functions, we need to work with a particle number $N$ for which both $\Phi_n$ and $\Phi_{\bar{n}_\uparrow,\bar{n}_\downarrow}$ can be constructed. That is possible for certain values of $N$ for spin-singlet states (i.e. when $\bar{n}_\uparrow=\bar{n}_\downarrow=\bar{n}/2$, but not for partially polarized states. We circumvent this problem by constructing wave functions with quasiholes in the spinless portion, e.g. in the $\Phi_n$ factor in Eq.~(5) of the main text.  In the thermodynamic limit, the addition of an order one number of quasiholes or quasiparticles does not alter the energy per particle, and therefore, we take this wave function to be an adequate representation of the partially polarized states. 

\begin{figure}[h]
	\includegraphics[width=0.95\columnwidth]{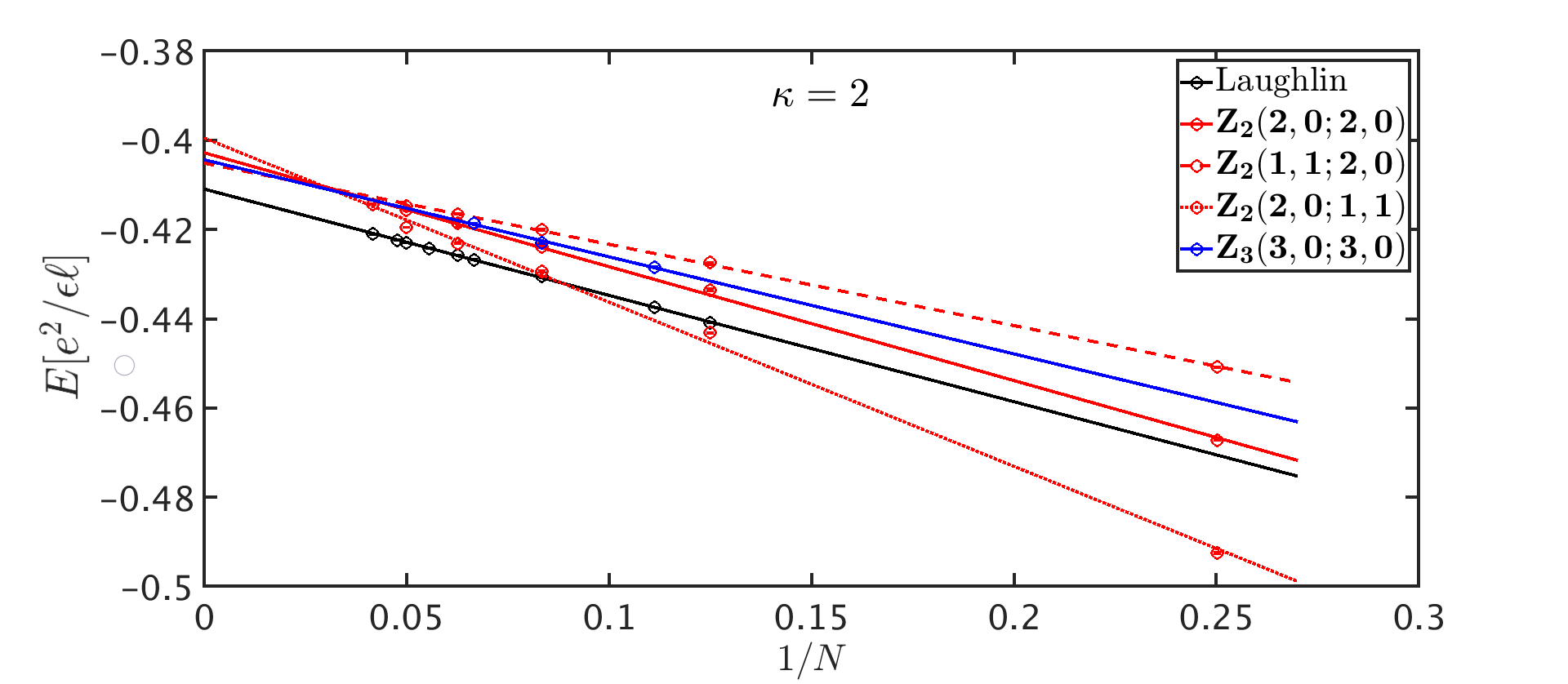}
		\includegraphics[width=0.95\columnwidth]{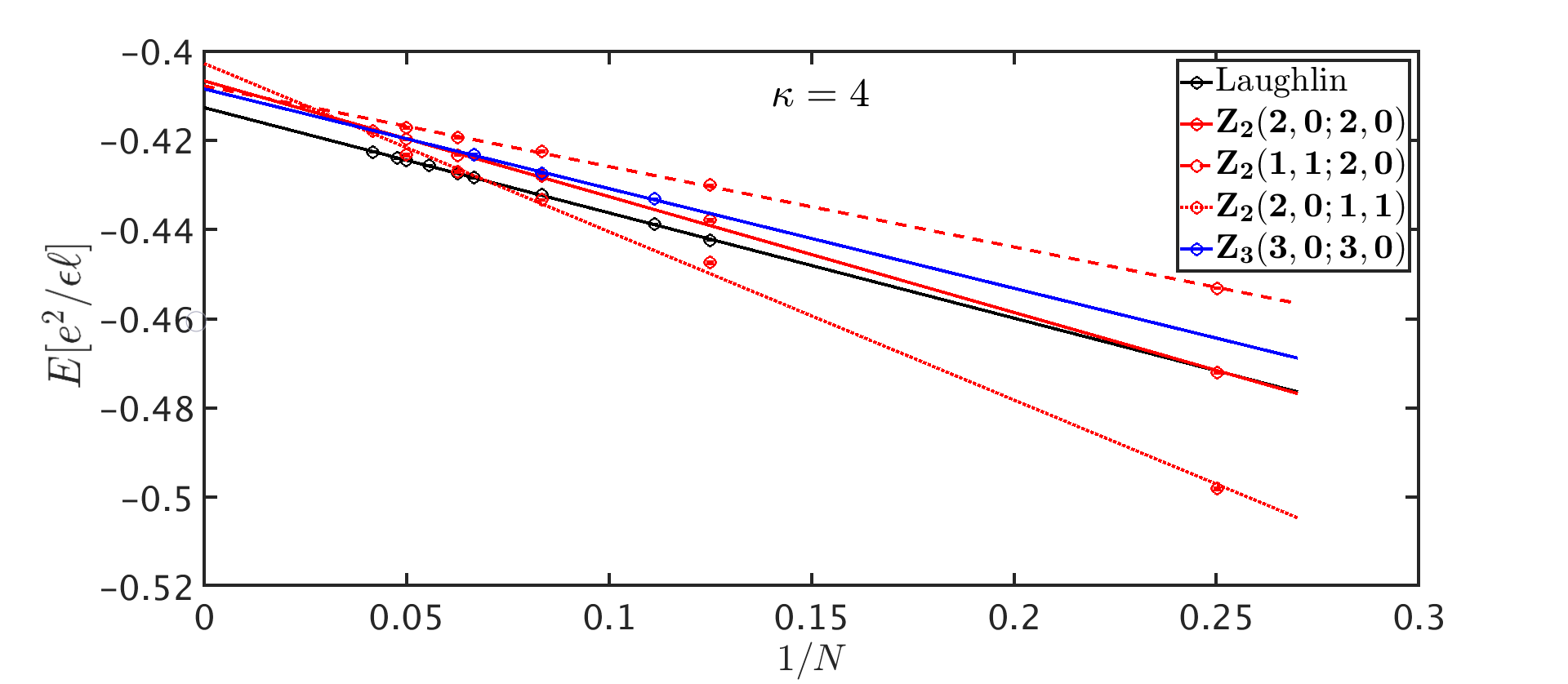}
		\includegraphics[width=0.95\columnwidth]{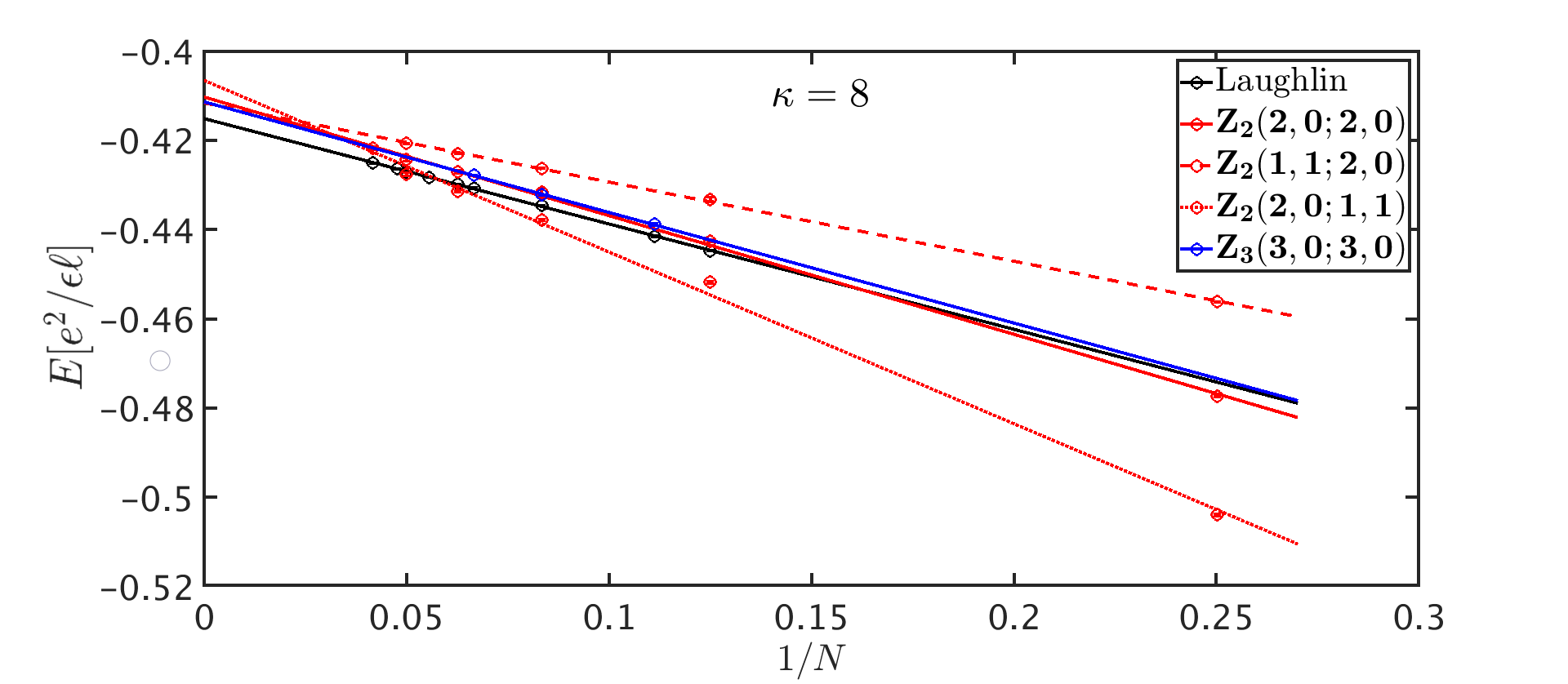}
	\caption{Thermodynamic extrapolations for the energies of several $\mathbb{Z}_n$ states at $\nu=1/3$ in the LLL. The LLM parameter is $\kappa=2$ (upper panel), $\kappa=4$ (middle panel), and $\kappa=8$ (lowest panel). The calculation is done for Coulomb interaction assuming zero width.}
	\label{fig: DMC_extrap_k2}
\end{figure}

\begin{figure}[h]
	\includegraphics[width=0.95\columnwidth]{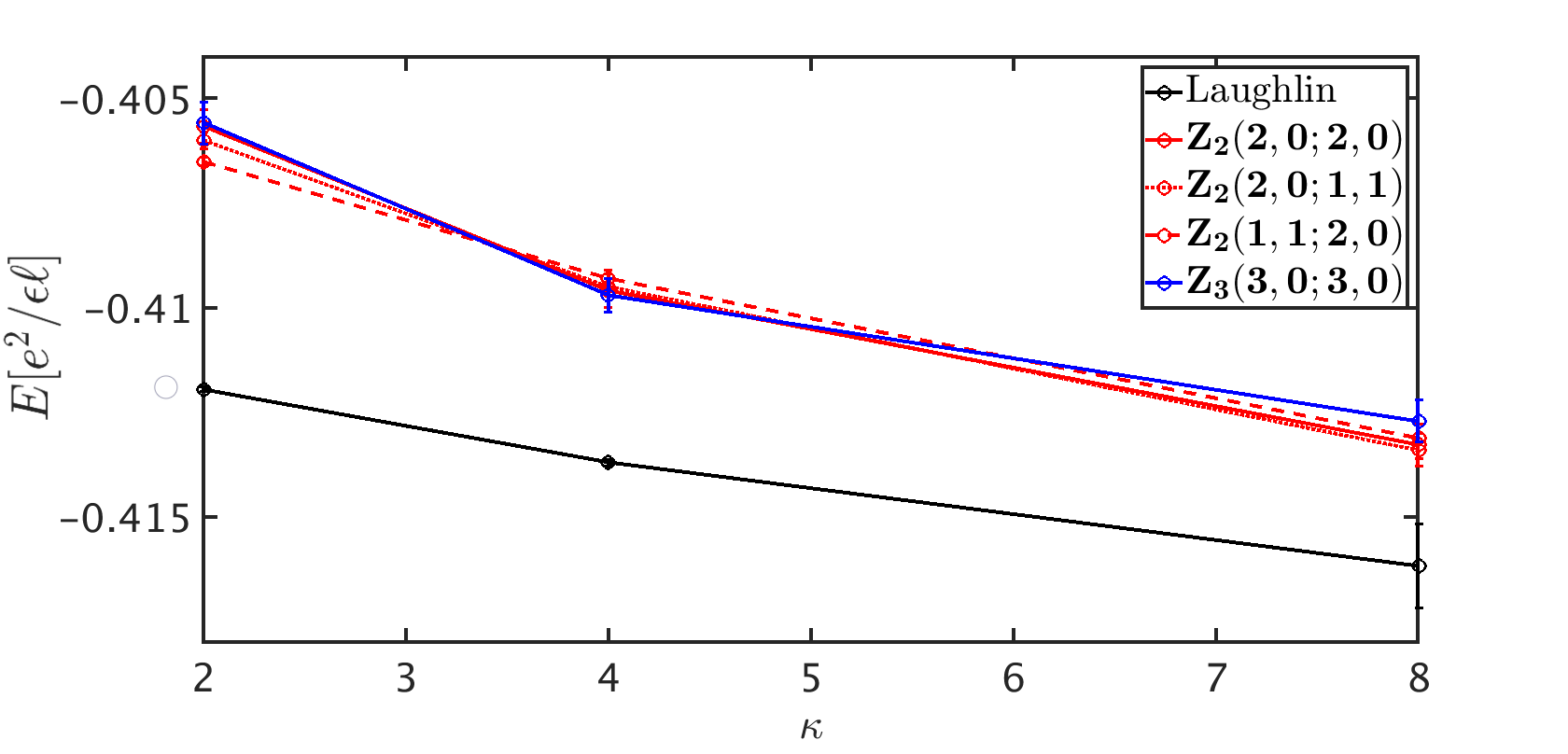}
	\caption{The thermodynamic energies of several $\mathbb{Z}_n$ states at $\nu=1/3$ in the LLL as a function of the LLM parameter $\kappa$. The calculation is done with a Coulomb interaction assuming zero width. The Laughlin state remains the ground state.}
	\label{DMC_all}
\end{figure}

We present in Fig.~\ref{fig: DMC_extrap_k2} energies from our fixed phase DMC calculations for $\nu=1/3$ in the LLL for several values of the LLM parameter $\kappa$. The thermodynamic energies are shown in Fig.~\ref{DMC_all}. Even though the energy differences are reduced with increasing $\kappa$, the Laughlin 1/3 state comfortably remains the ground state. 

\end{document}